\documentstyle[aaspp4]{article}

\newcommand\MSUNYR{\rm M_{\odot}\,yr^{-1}}
\newcommand\MSUN{\rm M_{\odot}}
\newcommand\LSUN{\rm L_{\odot}}
\newcommand\RSUN{\rm R_{\odot}}
\newcommand\Mdot{ \dot{M}}
\newcommand\etal{{\it et al}. }
\newcommand\be {\begin{equation}}
\newcommand\en{\end{equation}}
\newcommand\cm{\rm cm}
\newcommand\din{\rm din}
\newcommand\gr{\rm g}
\newcommand\seg{\rm s}
\newcommand\AU{\rm AU}
\newcommand\yrs{\rm yrs}
\newcommand\K{\rm K}

\slugcomment{Accepted by {\it The Astrophysical Journal}}

\begin{document}

\title
{ACCRETION DISKS AROUND YOUNG OBJECTS.I.
 THE DETAILED VERTICAL 
STRUCTURE} 

\author{Paola D'Alessio \altaffilmark{1}, Jorge Cant\'o  \altaffilmark{1}, Nuria Calvet \altaffilmark{2,3}, and
Susana Lizano \altaffilmark{1}}
 
\altaffiltext{1}{Instituto de Astronom\'\i a, Universidad Nacional Aut\'onoma de M\'exico,
A.P. 70-264, 04510 M\'exico D.F., M\'exico; Electronic mail: dalessio@astroscu.unam.mx, lizano@astrosmo.unam.mx}
\altaffiltext{2}{Harvard-Smithsonian Center for Astrophysics, 60 Garden St., Cambridge, MA 02138;
Electronic mail: ncalvet@medea.harvard.edu}
\altaffiltext{3}{On leave from Centro de Investigaciones de Astronomia, M\'erida, Venezuela}

\begin{abstract}
We discuss the properties of an accretion  disk around 
a star with parameters
typical of classical T Tauri stars (CTTS), and 
with the average accretion rate for these
disks. 
 The disk is assumed steady and geometrically thin. The turbulent viscosity 
coefficient is expressed using the $\alpha$ prescription  and the main 
heating mechanisms considered are viscous dissipation and 
irradiation by the central star. The energy is transported 
by radiation, turbulent conduction and convection.
 We find that irradiation from the central star 
is the main heating agent of the
disk, except in the innermost regions,  $R < 2 \ \AU$. The 
irradiation
increases  the temperature of the outer disk relative to the
purely viscous case. As a consequence, 
the outer disk ($R > 5 \ \AU$) becomes  less dense, optically 
thin and almost vertically isothermal, 
with a 
temperature distribution  $T \propto R^{-1/2}$. 
The decrease in surface density at the outer disk,  decreases the disk mass  
by a factor of $4$ respect to a purely viscous case. 
In addition,  irradiation 
tends to make the outer disk regions stable against gravitational 
instabilities.

\end{abstract}

\section{Introduction}

The co-planarity and circularity of planetary orbits in our Solar System
support the idea that it has been formed from a rotating
 ``protoplanetary disk'', 
 where energy is dissipated and momentum redistributed before the time of 
planet formation. 
Circumstellar disks are naturally formed 
during the collapse of a cloud fragment with non-zero
angular momentum (\cite{CM81}; \cite{TSC84}).
Disks produce observational signatures 
in the spectral energy distributions (SED) of young stellar objects (YSO)  in 
the form of 
excesses of emission at UV, IR and mm wavelengths.

During the  early evolutive phase of a forming star, 
 the disk-star system 
is still  surrounded by an infalling envelope, which dominates 
the far IR SED. At longer wavelengths 
the infalling envelope becomes optically thin, and the 
SED is dominated by emission from the disk (\cite{KCH93}; \cite{CHKW};   
\cite{HCB}).
Radiation from the infalling envelope is 
 one of the most important 
heating mechanisms of the disk
 in this early phase and has to be included to understand  the long wavelength 
SEDs of embedded 
stars 
 (e.g. \cite{butner91}, 1994; DCH97)

There can be a transition phase in which the star and the 
disk are surrounded by 
a tenuous dusty envelope which scatters stellar radiation 
onto the disk (\cite{natta93}), heating its outer regions. 
But finally, in a more evolved phase, 
it is expected that the material surrounding
the disk-star system becomes negligible (e.g. \cite{SAL87}).

Models for accretion disks (\cite{LBP}) 
 or flattened disks reprocessing stellar radiation (\cite{ALS87})
predict  steeper SEDs than those 
observed in Classical T Tauri stars.
Kenyon \& Hartmann (1987, hereafter KH) have proposed that a  disk 
in vertical hydrostatic equilibrium,  
with well mixed gas and dust, is flared and 
intercept more stellar radiation than a flat disk. 
This model predicts a SED more similar to the observed ones, but it is 
restricted to wavelengths $\lambda < 100 \ \ \mu {\rm m}$ because of their  assumption that the 
disk is 
optically thick.

In this paper,  we explore the evolved phase  in which 
the disk-star system is surrounded by a negligible amount of 
dust, 
 adopting the basic idea behind 
the KH model, i.e. gas and dust in the disk are  well mixed and thermally 
coupled. 
The disk receives radiation directly 
from the central star and it is also heated 
by viscous dissipation, cosmic rays and radioactive decay. 
We calculate the detailed vertical structure of 
such a disk taking into account several transport mechanisms: turbulent flux, 
radiation and convection.  
The disk interior and atmosphere are not arbitrarily separated.
The complete structure, from the midplane to the surface, is calculated with the 
same set of equations,  written in a general way 
to treat both optically thick and thin regions. 
We then present a detailed study of 
one case, with parameters typical of 
Classical T Tauri stars.  

In a forthcoming paper (paper II) 
we show results of the structure and physical properties
 for a wide range of parameters characterizing disks and central stars. 
Finally, motivated by
the large amount of observational information compiled in recent years,  
  we calculate  (paper III)
how the observational signatures of disks depend on their 
physical properties. 
Comparing models with observations 
of Classical T Tauri stars, we infer disks physical properties.

The structure of the present paper is as follows: in \S 2 
we present the assumptions and general description of the model; in 
\S 3 and \S 4 we give a detailed discussion of the heating of the disk and the 
energy transport; in \S 5 and \S 6 the equations and boundary conditions are 
written; in \S 7 a particular case is discussed, with 
parameters typical of Classical T Tauri disks; in \S 8 we compare 
the model with a non-irradiated accretion disk and discuss the effects of 
irradiation; in \S 9 the effect of the irradiation from an accretion ring 
is calculated  
and finally 
\S 10 presents a summary of the results.

\section{General description of the Model}

We assume that the disks are in {\it steady state}, thus, all their properties 
are 
time independent. We also assume that the disk is  {\it geometrically thin}, so 
the  
radial energy transport is neglected and 
the vertical and radial structures are treated as decoupled problems 
(\cite{LBP}; \cite{pringle81}).
The disk is assumed in {\it vertical hydrostatic equilibrium}. Since there 
can be turbulent and convective motions in the vertical direction, the 
assumption of hydrostatic equilibrium refers to the mean bulk motion of the gas 
and dust in the disk.
These assumptions are basic for the model   and 
cannot be relaxed without changing the 
disk structure equations and the numerical method used  to integrate them.
 Other additional assumptions, required 
to quantify the disk physical properties 
are that the  mass accretion rate $\Mdot$ is uniform through the disk  
 and  that the turbulent viscosity is described by 
the $\alpha$ {\it viscosity  prescription}. Therefore,  
the viscosity effective coefficient is given by  
 $\nu_t = \alpha \ c_s H$, where 
$c_s$ is the local sound speed, $H$ is the local scale height of the gas and 
$\alpha$ is a free parameter, which is 
assumed constant through the disk with the only constrain   
 that $\alpha \le 1$ 
 (\cite{SS73}). This prescription could be relaxed if $\Mdot(R)$ 
and $\nu_t(R,z)$ are given.

We consider 
that the disk is heated by viscous dissipation, 
radioactive decay, cosmic rays, and stellar irradiation. The first two
mechanisms generate and deposit energy at each height. Cosmic
rays and stellar irradiation penetrate the disk from the upper surface,
and interact with the material below.
We assume the disk is a mixture of gas and dust  and both components 
are thermally coupled 
and described by a unique temperature. 
The gas is locally heated or cooled by viscous dissipation, 
ionization by energetic particles,  
collisions with dust,  dust and gas radiation, convection, and turbulent 
conduction.  The dust grains are
 heated or cooled by radiation from the star and from the disk itself, 
and also by collisions with gas particles. 
 A more general description of 
the disk structure is required to calculate the disk vertical structure equations 
assuming that gas and dust have different temperatures. 
However, we assume that the 
collisional and/or radiative heating between gas and dust are 
efficient enough at every point in the disk and use only one mean temperature 
to describe both components.

Because the disk is steady, the energy absorbed is emitted and
transported outwards. 
 The energy is transported by:
(a) a {\it
turbulent flux}, computed self-consistently with the viscosity coefficient
used to describe the viscous energy dissipation; (b) {\it radiation},
through the first two moments of the transfer equation
; and (c) {\it convection},
described by the mixing length theory, taking into account that the convective 
elements
lose energy by radiation and turbulent flux. 

Given a central star, the important quantities  in describing a disk model are 
$\Mdot$, which gives the total energy flux generated by  viscous dissipation 
at each annulus in the disk and $\alpha$, which 
quantifies the fraction of viscous flux produced 
at each height. The disk structure also depends on the  
central star properties, i.e., 
its mass $M_*$, radius $R_*$ and effective temperature $T_*$.  
The mass $M_*$ is related to the depth of the stellar  gravitational 
potential, 
which is the last source of energy in 
 the accretion process. The stellar luminosity, related 
to 
  $T_*$ and $R_*$,  controls the amount of energy 
 introduced in the disk by stellar irradiation.

\begin{figure}
\plotone{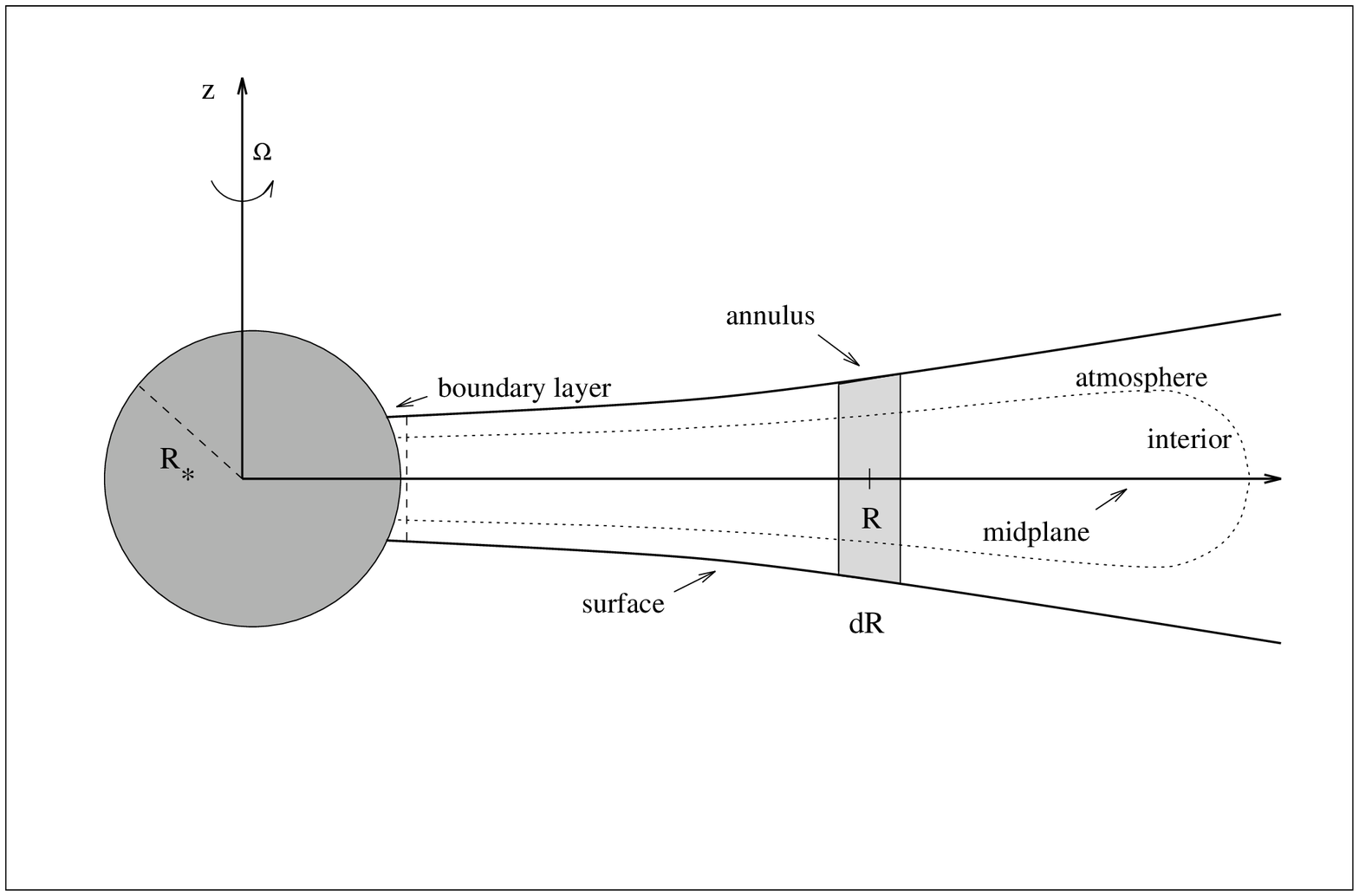}
\caption {Geometry of the disk. This plot shows schematically what is called 
annulus, surface, atmosphere, interior and midplane through the text. At a given 
distance $R$ from the central star, the disk vertical structure 
refers to the disk physical properties in the $z-$direction. 
The 
{\it atmosphere} refers to the optically thin part of the disk and the 
{\it interior} corresponds to the optically thick part; we indicate
schematically that the outer regions become optically thin.}
\label{fig_geomet}
\end{figure}

 The following sections give the 
physical and mathematical
description of these processes. The resultant set of equations
are solved with appropriate boundary conditions to get the
temperature and density at each radius and height of the disk. 
The disk geometry is shown schematically in Figure \ref{fig_geomet}.

\section {Heating sources}

The disk heating sources are viscous dissipation, ionization by 
energetic particles 
 and stellar radiation. In this section 
we describe in detail each of these mechanisms and give the corresponding 
rates to be used in the calculation of the vertical structure.

\subsection{Viscous dissipation}

The energy rate per unit volume locally generated by viscous stress is given by 
(\cite{pringle81}; \cite{FKR92}):

\be \Gamma_{vis} = {9 \over 4} \ \rho(z) \  \nu_t(z) \  \Omega(R)^2 = {9 \over 4} \alpha \  P(z) \  \Omega(R) \  
\ \ ,  
\label{eq_dfviscosa}
 \en

\noindent
 where $\rho$ is the mass density, 
$P$ is the gas pressure, $\Omega$ is the Keplerian 
angular velocity, and 
$\nu_t$ is the turbulent viscosity coefficient, given by $\nu_t=\alpha \ H \ c_s$.

 The total  energy per unit area produced by viscous dissipation  
in a given annulus at a distance  $R$ from the central star is thus, 

\be F_{vis}(z_\infty)= \int_0^{z_\infty} \ \Gamma_{vis} \ dz=   {3 G M_* \Mdot 
\over 8 \pi } R^{-3} \biggl
[ 1 - \biggl ( {R_* \over R} \biggr ) ^{1/2} \biggr ]  \ \ , 
\label{eq_fvis}  \en
where $z_\infty$ is the height of the disk, and
 $G$ is the gravitational constant.

\subsection{Ionization by energetic particles}

The main source of ionization by energetic particles are cosmic rays. 
However, Stepinski (1992) has shown that ionization due to energetic particles 
produced during 
the radioactive decay of $^{26} {\rm Al}$ can be more important in those regions 
where cosmic rays can not penetrate. 
We have thus considered both mechanisms and the resulting heating rate per unit 
volume is written as, 

\be  \Gamma_{ion}(\Sigma)  =  n_{H_2} \  [\Delta Q_{cos} \ \zeta_{cos} \ 
e^{-\Sigma/\lambda} +
\ \Delta Q_{rad} \  \zeta_{rad}(^{26}Al) ]  \ \ ,
\label{eq_heatcr}
\en

\noindent
where 
   $n_{H_2}$ is the number density of Hydrogen molecules,
  and $\lambda = 96 \ \gr/\cm^2 \ $ is the attenuation surface density scale  
 of cosmic rays. The ionization rates due to cosmic 
rays 
and radioactive decay are 
$ \ \zeta_{cos} = 10^{-17} \ \seg^{-1}$ (\cite{NU86}) 
 and $ \ 
\zeta_{rad}(^{26}Al) = 5.2 \times
10^{-19} \ \seg^{-1}$ (\cite{step92}),
 respectively, 
and $\Sigma$ is the vertical column density 
from the disk surface to a height $z$.
We have  taken the energy liberated in  heating in each cosmic ray ionization as
 $\Delta Q_{cos} \approx  20 \ {\rm eV}$, and for radioactive decay ionization,
   $\Delta Q_{rad} \approx 10 {\rm eV}$
 (\cite{GL78}).

Integrating  equation (\ref{eq_heatcr}) in $z$, 
the total energy input
 per unit area due to ionization by energetic particles 
 is given by:

\be
F_{ion}(z_\infty) = \int_0^{z_{\infty}} \ \Gamma_{ion} \ \  dz = {1 \over 2 m_H 
}  \{ \Delta Q_{cos} \ \zeta_{cos} \ \lambda \ [1 - e^{-\Sigma_\infty / 2 \lambda} ] + 
\Delta Q_{rad} \ \zeta_{rad} \ {\Sigma_\infty \over 2}  \}  \ \ , 
\label{eq_fcosm}
\en

\noindent
where $m_H$ is the mass of the hydrogen atom and  
$\Sigma_\infty$ is the total column density of the disk, 
\be
\Sigma_\infty = \int_{-z_\infty}^{z_\infty} \rho \ dz \ \ .
\label{eq_sigma}
\en

\subsection{Stellar irradiation}

Following  Calvet \etal (1991, hereafter CPMD), it is assumed
that  the stellar radiation intercepted by 
the disk consists of a parallel beam 
carrying a flux $F_{irr}$ per unit area of the disk surface,  
incident at an angle $ \theta_0 = \cos^{-1} \mu_0$ from the normal to the 
boundary of the disk atmosphere. A fraction  of this energy 
 is {\it scattered} creating a diffuse field 
with wavelengths around the characteristic 
wavelength of the stellar radiation, given by the 
stellar effective temperature $T_*$. The remaining fraction 
 of the incident radiation field is truly absorbed and remitted 
at wavelengths determined by the local temperature of the disk.
 As long as the characteristic temperature 
of the disk is lower than the effective temperature of the star, it can be 
assumed  
 that  no true emission 
of the disk occurs in the ``stellar'' frequency range and no scattering 
of the incident beam occurs in the ``disk'' frequency range.

The true absorption of radiation at the stellar frequency range  
is described through a Planck-type mean opacity, calculated evaluating the 
true absorption coefficient at the local temperature and pressure but  
using the Planck function evaluated at $T_*$ as the weighting function. 
 The resulting mean absorption coefficient 
is  given by:

\be
\kappa_P^*(T,P,T_*) = {\pi \over  \sigma_{R} T_*^4}\int_0^\infty 
\kappa_\nu(T,P)B_\nu(T_*) d \nu \ \ ,
\label{eq_kaps}
\en

\noindent
where $\kappa_\nu$ is the monochromatic true absorption coefficient, evaluated 
at the gas temperature $T$ and gas pressure $P$, and 
$\sigma_R$ is the Stefan-Boltzmann constant. 
The mean scattering coefficient  $\sigma_P^*(T,P,T_*)$,  is calculated 
with  
the same kind of average given by equation (\ref{eq_kaps}), substituting 
$\kappa_\nu$ by $\sigma_\nu$, the monochromatic scattering coefficient.
The mean extinction coefficient is 
 $\chi_P^*(T,P,T_*)=\kappa_P^*+\sigma_P^*$. 
Since we 
assume that gas and dust 
are well mixed in the disk at every height, the absorption and scattering coefficients include the contribution of dust, 
unless it is sublimated in 
regions with a temperature higher than the sublimation temperature.
If the dust has settled down in the midplane, the upper disk atmosphere would 
have a lower 
 opacity to the stellar radiation than what we are assuming 
here.

 The impinging stellar radiation field, which propagates 
in the direction defined by $\mu_0$, 
reaches the height $z$  with a flux given by:

\be
F_{i}(z) = - F_{irr} \ e^{-\tau_s/\mu_0} \ \ , 
\label{eq_dirflux}
\en

\noindent
where the optical depth $\tau_s$ is:

\be
\tau_s = \int_z^{z_\infty} \chi_P^* \  \rho \  dz \ \ .
\label{eq_taus}
\en 

\noindent
and the minus sign reflects the fact that this flux is going inward. The 
corresponding  mean intensity is given by:
\be
J_{i}(z) = {F_{irr} \over 4 \pi \mu_0} \ e^{-\tau_s/\mu_0} \ \ ,
\label{eq_js}
\en

The fraction of the incident radiation which 
is scattered is $s=\sigma_P^*/\chi_P^*$ 
and the fraction absorbed is $\kappa_P^*/\chi_P^*=1 - s=a$.
For simplicity, it is assumed that $a$ and $s$ are constant 
in the region where the stellar radiation is mainly absorbed 
(both fractions are evaluated at T and P such that $\tau_s=2/3$). 
This assumption will be checked {\it a posteriori}.

Following CPMD, the zero-order moment of the  transfer equation for 
the diffuse radiation field can be written as: 

\be
{dF_s \over d\tau_s} = a 4 \pi J_s - s \ {F_{irr}  \over \mu_0} 
e^{-\tau_s/\mu_0} \ \ ,
\label{eq_atransps}
\en
where $F_s$ and $J_s$ are the flux and mean intensity
 of the diffuse stellar radiation field.

The first moment of the transfer equation, assuming the 
Eddington approximation (i.e., that the diffuse field is isotropic),
 is given by (\cite{M78}):
\be
{dJ_s \over d\tau_s} = 3 {F_s \over 4 \pi}
\label{eq_btransps}
\en

From equations (\ref{eq_atransps}) and (\ref{eq_btransps}), 
  a second 
order differential equation for $J_s$ can be constructed.
Given the boundary condition corresponding to an isotropic radiation field, i.e.
 $J_s= F_s/2 \pi$, and assuming that $s$ and $a$ are independent on $\tau_s$,
 the diffuse radiation field is described by: 

\be
J_s = {s \ F_{irr} (2 + 3 \mu_0) \over 4 \pi [1 + (2 g/3)](1 - g^2 \mu_0^2) }
e^{-g \tau_s} - {3 \mu_0 \ s \ F_{irr} \over 4 \pi (1-g^2 \mu_0^2) } 
e^{-\tau_s/\mu_0},
\label{eq_jnuria}
\en

\be
F_s = -{g \ s \ F_{irr} (2 + 3 \mu_0) \over 3  [1 + (2 g/3)](1 - g^2 \mu_0^2) }
e^{-g \tau_s} + { s \ F_{irr} \over  (1-g^2 \mu_0^2) } e^{-\tau_s/\mu_0},
\label{eq_fnuria}
\en
 
\noindent
where $g=\sqrt{3a}$.

Finally, from  equations (\ref{eq_dirflux}) and (\ref{eq_atransps}), the disk 
heating rate per unit volume due to  stellar radiation can be written as:  

\be
{\Gamma_{irr}}= \chi_P^* \rho \biggl [4 \pi a J_s + (1-s) {F_{irr} \over \mu_0} 
e^{-\tau_s/\mu_0} \biggr ] = 4 \pi \kappa_P^* \rho [J_s+J_i] \ \ .
\label{eq_firrad}
\en

The irradiation flux $F_{irr}$ is calculated as described by KH, 
with the disk height and shape 
calculated self-consistently, under the assumption of gas and dust well mixed and 
thermally coupled.

\section{Energy transport}

 In different regions of a 
 given disk, 
 different mechanisms could be responsible of 
the energy transport. With the  aim of making 
a reliable model to cover  a wide range 
of disk and stellar parameters, we have considered 
that the energy in the disk can be transported by radiation, convection and 
turbulence. In the following we describe each of these mechanisms.

\subsection{Radiative transport}

The  disk's own radiation 
is characteristic of the disk local temperature,  so it 
corresponds to lower frequencies respect to 
those characteristic of the stellar radiation.
 The radiative transport at these frequencies    
 is described through the first two moments of
the transfer equation integrated in frequency, using 
the Eddington approximation to close the system (\cite{M78}).
Instead of mean opacities averaged taking as   
weighting functions the frequency dependent 
mean intensity and flux ($\kappa_J$ and $\chi_F$, in Mihalas notation), the  
 Planck and Rosseland mean opacities ($\kappa_P$ and $\chi_R$) have been used. 
These 
mean opacities  
are computed with 
a consistent set of monochromatic opacities (see CPMD and 
\cite{D96}, for details).  

The first two moments of the transfer equation in a frequency  range characteristic 
of the disk physical conditions, 
 can be written as:

\be
{dF_d \over dz} = \Lambda_d-\Gamma_d=
4 \pi \kappa_P \  \rho \  \biggl [ {\sigma_R T^4 \over \pi} - J_d \biggr] \ \ , 
\label{eq_dfddz}
\en

\be
{dJ_d \over dz} = -3 \chi_R \  \rho \  {F_d \over 4 \pi} \ \ .
\label{eq_djddz}
\en

In equation (\ref{eq_dfddz}), 
$\Lambda_d$ represents 
the radiative cooling rate per unit volume, and 
$\Gamma_d$,  
the heating rate per unit volume due to disk radiation (i.e. in the disk 
frequency range).

Finally, the energy per unit area per time transported by radiation is given by 
the direct stellar flux, the diffuse stellar flux and the disk radiative flux, 
that is, 

\be
F_{rad} = F_i+F_s+F_d \ \ .
\en

\subsection{ Turbulent flux}

The turbulent elements 
responsible for the disk viscosity are also transporting energy
in the vertical direction (\cite{rudiger}).
Here we have assumed that the turbulence in the disk  
 is not generated by convection but by other mechanism (e.g. 
instabilities associated with  magnetic fields, see \cite{BH91}, 1992; \cite{HB91}, 1992). 
So convection and turbulent fluxes are treated as different 
energy transport mechanisms, which can be both present in a given
region of the disk.

An adiabatic turbulent element moving from a hot region to a colder one in a 
superadiabatic medium has, through its 
trajectory, an excess of energy with respect to the surrounding medium. 
It loses part of its energy doing work 
on the surrounding gas   
 and the rest of the energy  is liberated 
when it finally dissolves, mixing with the medium.
If the medium is subadiabatic, the rising adiabatic elements have less energy 
with respect to the ambient gas and cools the upper and cooler layers.    
Assuming there is no net vertical mean motion in the disk, there are 
the same number of elements rising and falling at each height, so 
   the net energy interchanged between the turbulent elements and 
the medium along their path is zero. Only the excess or 
deficit of energy of the elements and the medium,  
 just before dissolving,  contributes
to the net flux. If the turbulent elements 
are adiabatic,  the turbulent energy flux is proportional to
the entropy gradient, and  
following R\"udiger \etal (1988), can be written as:

\be F_{turb}= -\rho T { \nu_t \over P_r} {dS \over dz} = {\alpha \  (P+P_{rad}) 
\over \Omega  P_r}
  g_z \biggl [ {\nabla \over \nabla_A} -1  \biggr ]  \ \ ,
\label{eq_fturb}  \en

\noindent
where $S$ is the specific entropy,  
 $\nabla$ is the gradient $(d \ ln T/d \ ln P)$ of the medium, 
$\nabla_A$ is the adiabatic gradient (i.e. $\nabla$ evaluated at constant 
entropy) and $P_{rad}$ is the radiation  pressure. The 
Prandtl number $P_r$ is  
 given by the 
ratio of the efficiency of momentum and energy transport by 
turbulent elements, i.e.,  $P_r = 
\nu_t/\chi_t$, where $\nu_t$ is the turbulent viscosity coefficient and
 $\chi_t$ is the turbulent conductivity coefficient. Finally,  
$g_z$ is
the z-component of the stellar gravity, given by:

\be g_z= \Omega^2 \ {z \over [1 + (z/R)^2]^{3/2} }   \ \ .
\label{eq_gravity}
\en 

\subsection{ Convection}
The energy transport by convection is included 
through the mixing length  theory 
(e.g. \cite{CG68}), with a convective efficiency computed assuming
that the eddies loose energy by turbulent flux and by  
 radiation. The radiative losses are calculated   
estimating the optical depth of the elements (\cite{M78}).
The temperature gradient of the medium is calculated given the total flux of
energy that has to be transported at each point.  
In order to calculate the energy transport by convection 
in both optically thin and thick convective regions 
of the disk and
also to take into account that part of 
the total energy flux $F$ is transported by turbulent
conduction, instead of using 
the usual definition of radiative gradient $\nabla_R$ based on the
diffusion approximation, we introduce the gradient $\nabla_{RC}$.
This is the gradient $(d \ lnT /  d \ ln P)$ of the medium when convection is 
absent 
and can be written as:

\be \nabla_{RC} \equiv  {F-F_{rad} \over A_{turb} }+ \nabla_A  \ \ , \en

\noindent
 where  
$A_{turb}  \equiv  \alpha P \ g_z  / \Omega P_r  \nabla_A$ and we have used 
equation (\ref{eq_fturb}).

 It is assumed the  disk is unstable to convection in those regions 
 where 
 $\nabla_{RC} > \nabla_A$.
When this condition holds  the true gradient of the medium is given by:

\be
\nabla= (1- \zeta) \ \nabla_{RC} + \zeta \ \nabla_A \ \ ,
\label{eq_nabla}
\en
where $\zeta$ is a convective efficiency 
 and depends on  
 the mixing length $\Lambda$,  a free 
parameter of this convection theory (see \cite{D96}  for details).
On the other hand, when $\nabla_{RC} < \nabla_A$ the medium is stable against 
convection and  energy is transported only by radiation and turbulent flux. 
In these regions  
 $\nabla=\nabla_{RC}$.

 The quantity $\nabla$ contains the information of the efficiency of 
the different 
energy transport mechanisms at each point in the disk.  Given $\nabla$, 
 the differential equation for the kinetic temperature  can be 
written as:

\be {dT \over dz} = -\nabla  {T \over P} g_z \rho      \ \ , 
\label{eq_dtdz} \en

\noindent
 which describes 
the temperature structure required by the energy  transport mechanisms 
and the assumption of steady state.

 The adiabatic gradient $\nabla_A$ and other thermodynamical quantities 
used in this work
have been calculated in the way described by Vardya (1965) and Mihalas (1967).

\section{Hydrostatic equilibrium}
\label{sec_hidro}

The disk is assumed in vertical hydrostatic equilibrium, 

\be {dP \over dz} \ 
 = - \rho g_z  - {dP_{rad} \over dz}, 
\label{eq_hyd}
\en
where the radiation pressure  $P_{rad}$ is 
 included because it 
could be important for disks with very high accretion rates 
 or  in low density regions, like the upper atmosphere.
The radiation pressure gradient is 
given by (see \cite{M78}, pag. 170):

\be
{dP_{rad} \over dz} = {\rho \over c} \ \int_0^\infty \chi_\nu F_\nu d\nu \ \  , 
\en

\noindent
where $\chi_\nu$ is the monochromatic opacity coefficient, $F_\nu$ is the 
monochromatic radiative flux and $c$ is the speed of light. Using the
 mean opacities and the radiative energy fluxes defined in  sections 3.3 and  
4.1, the radiation pressure gradient can be approximated as:

\be
{dP_{rad} \over dz} \approx {\rho \over c} [ \chi_R F_{d} + \chi_R^*(F_s+F_i)] =
 {\rho \over c} [ \chi_R (F_{rad}-F_s-F_i) + \chi_R^*(F_s+F_i)] \ \ ,
\label{eq_radpress}
\en

\noindent
where 
 $\chi_R$ is the Rosseland mean opacity, $\chi_R^*$ is a 
Rosseland type mean opacity calculated  using  
 the derivative 
of the Planck function evaluated at the stellar 
temperature as the weighting function.

\section{Equations}

From the previous discussion, the set of differential equations  
that describe the disk vertical structure is:

\be
{dF_{rad} \over dz}=\Lambda_d - \Gamma_d-\Gamma_{irr} \ \ ,
\label{eq1}
\en

\be
{dJ_d \over dz} = - {3 \over 4 \pi} \ \chi_R(P,T) \rho {(F_{rad}-F_s-F_i)} \ \ ,
\label{eq2}
\en

\be
{dF \over dz} = \Gamma_{vis} + \Gamma_{ion} \ \ ,
\label{eq3}
\en

\be
{dT \over dz} = - \nabla(F_{rad},F,T,P) \ {T \over P} \ g_z \ \rho \ \ ,
\label{eq4}
\en

\be {dP_g \over dz} \ 
 = - \rho g_z  - {dP_{rad} \over dz} \ \ .
\label{eq5}
\en

 These equations 
 are solved subject to the following boundary conditions.

 At $z=z_\infty $:

\be P_g= P_\infty,
\label{eq_pcondsup} \en

\noindent
where $P_\infty$ is 
 a fixed small and arbitrary
 value of the gas pressure at $z=z_\infty$ (we 
adopt 
 $P_\infty=10^{-9} \ \din \ \cm^{-2}$). The total energy flux is: 

\be F \ = F_{rad} = \  \ (F_{vis} + F_{ion})|_{z_\infty}. 
\label{eq_fcondsup}\en

The net flux produced by viscous dissipation at $z=z_\infty$
 is $F_{vis}(z_\infty)$, given by equation (\ref{eq_fvis}),
 and the net flux produced by energetic 
particles ionization at $z=z_\infty$ is $F_{ion}(z_\infty)$ from equation 
(\ref{eq_heatcr}). 
The turbulent and convective fluxes are
zero, since by definition the eddies
cannot get out from the disk surface. 
 The stellar flux  going into the disk at $z_\infty$  is $F_{irr}$. 
This flux is reprocessed by the disk and emerges from its surface 
at a frequency range characteristic of its photospheric temperature. 
From the steady state assumption, the incident stellar flux is equal 
to the emergent stellar reprocessed flux, both integrated in frequency. 
Thus, the net flux associated with the stellar irradiation 
is zero at $z_\infty$ and it does not contribute to equation 
 (\ref{eq_fcondsup}).
  All the energy 
flux 
 emerging from the surface of
the disk is transported by radiation, i.e. 
$F_{rad}(z_\infty)=F(z_\infty)$. Finally,  

\be J_d (z_\infty) \  = J_\infty
=  {1 \over 2 \pi}  [F_{vis} + F_{ion}+ F_{irr}-F_s]|_{z_\infty}, 
\label{eq_jcondsup}
 \en
\noindent
where 
the irradiation flux $F_{irr}$ is the stellar flux 
intercepted by the disk surface. 
The mean intensity $J_\infty$ at the disk frequency evaluated at the disk 
surface, is calculated with the two stream approximation,  
subtracting  the contribution of the diffuse radiation field 
to the total stellar flux because 
it is scattered in the stellar frequency range.

 At $z \ = \ 0 $ : 

 \be F \ = F_{rad} =  \ 0, 
\label{eq_condsim}
\en
\noindent
i.e., 
at the disk mid-plane ($z=0$) all the energy fluxes
are zero because of 
 the reflection symmetry.

This is a two boundary problem. The disk heigh $z_\infty$ is an unknown boundary 
and has
to be determined. 
Using a $4^{th}$ order Runge-Kutta, we solve a reduced set of equations
corresponding to the diffusion approximation. This result is used as an 
initial guess for a relaxation method (\cite{press89}) 
 to integrate the full 
set of equations discussed above, taking $z_\infty$ as an eigenvalue.

We assume in this work that the disk is not surrounded by any
substantial external material (i.e. an envelope or a wind). With this
assumption, we choose a small value for $P_\infty$ such that any
material above it has no effect on the vertical structure; with this
value of $P_\infty$, we find that the outer disk regions flare.
However, we also find that for higher values of $P_\infty$, there can
be an outer region where the disk height has a maximum and then
decreases with $R$, so that the slope of the disk surface becomes
negative and the the computed irradiation flux becomes zero. But to
have higher values of $P_\infty$, the disk should be in pressure
equilibrium with some surrounding material. In this case, this material
would be heated by the central star and could transport scattered and
reprocessed stellar radiation to the deeper disk regions, which cannot
``see'' the star directly (Natta 1993; D'Alessio, Calvet, \& Hartmann
1997). If this quasi-isotropic heating was included, the outer disk
regions would flare (D'Alessio 1996; D'Alessio, Calvet, \& Hartmann
1997).

\clearpage
\section {Results and discussion}

For a central star of given  $M_*, R_*$ and $T_*$, a disk mass
accretion rate $\Mdot$ and a viscosity parameter $\alpha$,
 the density and temperature structure
  can be calculated by integrating equations
 (\ref{eq1}) to (\ref{eq5}). In this paper, we have chosen, as an example, 
 a set of parameters that are typical of T Tauri stars in Taurus: 
$\Mdot=10^{-8} \ \MSUNYR$ (\cite{VBJ93};
\cite{gull97}), $T_*=4000 \ \K$, $M_*=0.5 \ \MSUN$, and $R_*=2 \ \RSUN$ 
(\cite{KH95}). The assumed mass accretion rate  is lower than the critical mass accretion rate that leads to thermal instability (see \cite{KM93}, \cite{BL94}), justifying the steady state 
assumption of the present model. 
The last parameter is the viscosity coefficient $\alpha$. The value of
$\alpha$ is crucial for modelling of the vertical structure of accretion
disks, but is not known observationally. For the model discussed in this paper we
chose $\alpha=0.01$. This value is 
 consistent 
with angular momentum transport by turbulence initiated by the BH 
magnetohydrodynamical instability (\cite{BH91}, \cite{HB91},\cite{BH92}, 
\cite{HB92}, \cite{gammie}). 
Also,  as shown by D'Alessio (1996), T Tauri's observed SEDs
 at
$mm$ wavelengths can be fitted by irradiated disk models 
with different combinations of $\Mdot$ and $\alpha$. In particular, 
for $\Mdot=10^{-8} \ \MSUNYR$, the viscosity parameter which leads  
to a flux and slope around $\lambda=2.7 \ mm$ consistent with 
observations (\cite{dutrey96})  is 
$\alpha=0.01$.
 In two forthcoming papers, we will present (1) a grid of models 
to illustrate the effect of different parameters on the SEDs, and (2) a
detailed comparison between models and observations, to determine the
range of parameter space that yields the best fit to observed SEDs.

\subsection{Optical Depths}

Figure \ref{fig_taus} shows the radial distribution 
of the Rosseland $\tau_R$, the Planck $\tau_P$ 
and the ``stellar'' $\tau_s/\mu_0(R,z_\infty)$  mean total optical depths, integrated from 
the surface to the midplane of the disk.  
In this model $\tau_R >1$ for $R < 20 \ \AU$. For larger radii 
the disk becomes optically thin, decreasing $\tau_R$ with radius.
 The disk is optically thick to 
the stellar radiation ($\tau_s/\mu_0 >> 1$)  although it can be optically thin to its own 
radiation ($\tau_R < 1$). 
This is a consequence of the 
shorter wavelength characterizing the stellar radiation (CPMD) and 
the fact that it penetrates in a slanted angle  
(CPMD, Malbet \& Bertout 1991).
The Rosseland optical depth decreases for small radii ($R < 0.1 \ \AU$) because 
the sublimation of dust, which is the main opacity source 
in the disk. 
At these radii the disk temperature
becomes similar to the sublimation temperature, given approximately 
by  $ T_{sub} \approx 1800-2000 \ \K$.

The difference between the Planck and Rosseland mean optical depths reflects
the strong frequency dependence of the monochromatic opacities used to calculate 
the means. 
While $\tau_R$ controls the radiative transport in optically thick regions, 
$\tau_P$ controls the radiative cooling in optically thin regions.
Because radiation is the most important mechanism in transporting energy 
inside the disk
 (see below)
the characteristic temperatures are highly dependent on
the radial distribution of $\tau_R$ and $\tau_P$

\subsection{Characteristic Temperatures}

Figure \ref{fig_temp} shows the temperature at the disk 
midplane $T_c$ 
 and the photospheric temperature, (where 
$\tau_R=2/3$) $T_{phot}$.
For small radii ($R \lesssim 0.5 \ \AU$ in this model) flaring is not important 
and $T_{phot} \sim R^{-3/4}$ as in an irradiated flat disk. For larger radii,
 the disk curvature becomes important, increasing 
the amount of stellar flux intercepted and flattening 
the distribution of photospheric temperature, giving $T_{phot} \sim R^{-1/2}$.
This result is consistent with  the temperature distribution  
found by Miyake \& Nakawa (1995), for an optically thick flared disk, with 
gas and dust well mixed. 
 We only show the photospheric temperature for regions where
$\tau_R < 2/3$; outside, it is not defined.
A simple analytical approximation 
gives $T_{phot} \sim R^{-3/7}$ for the region where 
the disk flaring becomes important, assuming 
that the stellar radiation is deposited at a height 
$z_s \propto  H$, where $H$ is the gas pressure scale height 
(see \cite{CDL95}).
In this paper, the disk surface is assumed at constant pressure instead of 
proportional to the scale height, which results in a different radial dependence of the irradiation flux.
We will show below that $z_\infty$ is not proportional to $H$.

For   $R \lesssim  2 \ \AU$, 
viscous dissipation is the main energy source  
 of the regions near the disk midplane.
   At larger radii, $T_c$ 
 becomes higher than the central temperature 
 of a disk heated only by viscous dissipation, 
 reflecting the importance of irradiation 
 as a heating source. 
 The contrast 
 between $T_c$ and $T_{phot}$ 
 depends on  
$\tau_R$, 
 since this optical depth controls the vertical 
 temperature gradient 
 required to transport the energy released at every height.

\begin{figure}
\plotone{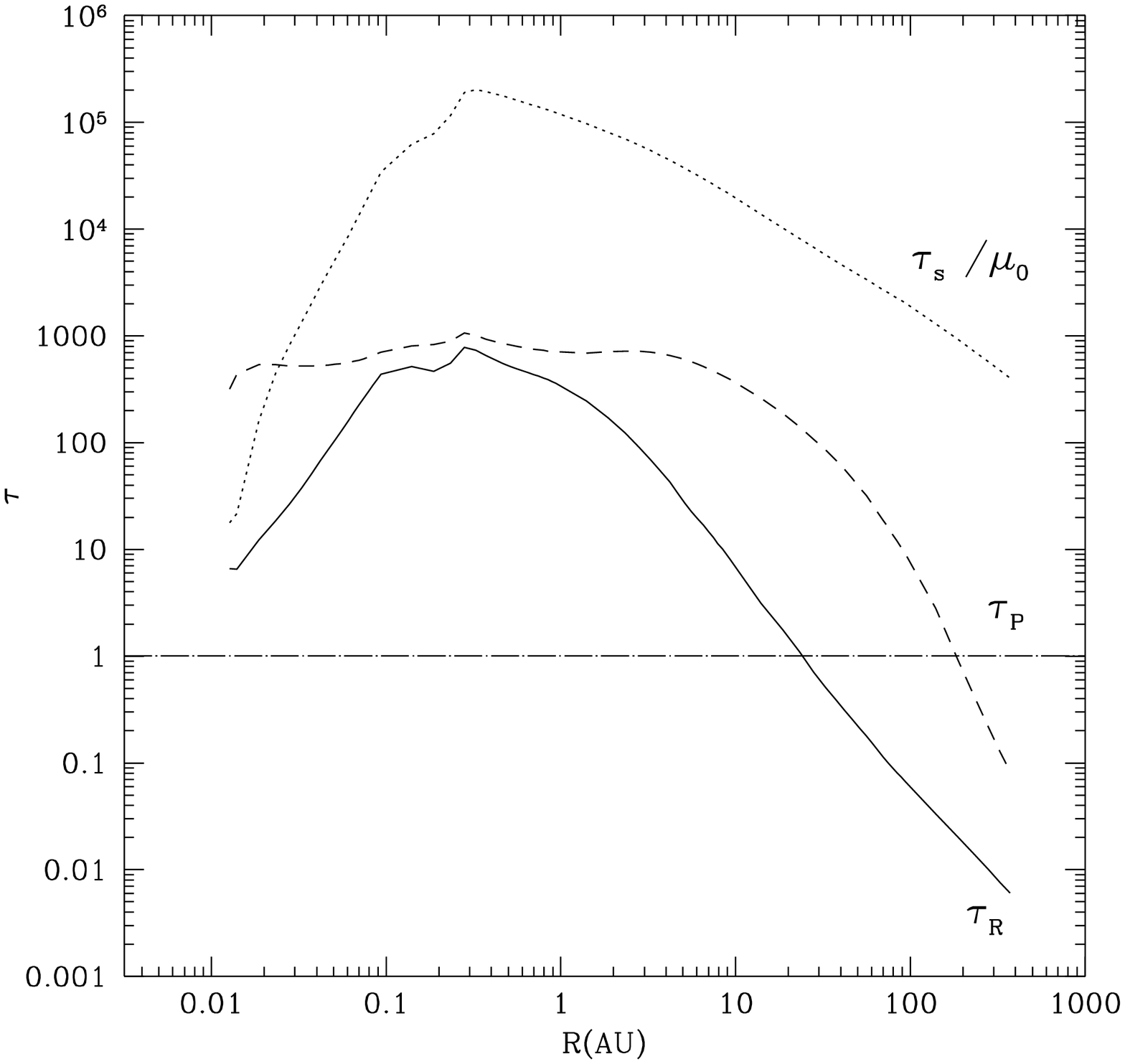}
\caption {Radial distribution of different mean optical depths, 
integrated 
 from the disk 
midplane to its surface. 
The disk parameters are $\Mdot=10^{-8} \ \MSUNYR$ and $\alpha=0.01$, the central star parameters are $M_*=0.5 \ \MSUN$, $R_*=2 \ \RSUN$ and $T_*=4000 \ K$. 
The vertical structure was calculates from: $R=0.0127 \ \AU$ to $373 \ \AU$.
The plotted optical depths are: Rosseland mean $\tau_R$ (solid line), Planck 
mean $\tau_P$ (dashed line) and the mean optical depth at the stellar 
frequency range $\tau_s/\mu_0(R,z_\infty)$ (dotted line). The horizontal line corresponds to 
$\tau=1$. }
\label{fig_taus}
\end{figure}

\begin{figure}
\plotone{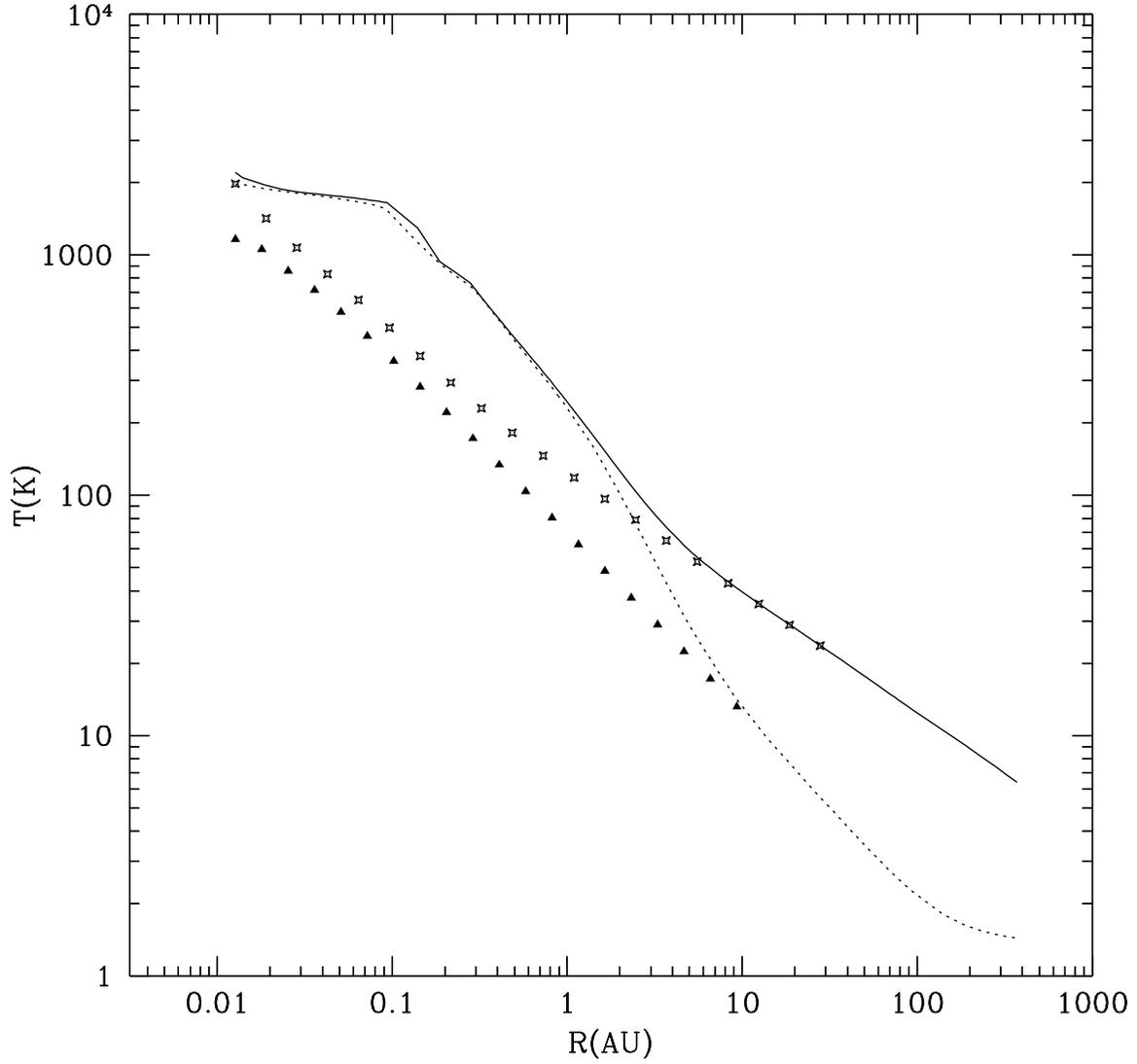}
\caption {Radial distribution of the midplane $T_c$ and photospheric $T_{phot}$
 temperatures 
for an irradiated and a non-irradiated disk model.
The plotted temperatures are: $T_c$ for the irradiated disk (solid line), $T_c$ 
for 
the non-irradiated disk (dotted line), $T_{phot}$ for the irradiated disk 
(stars) and $T_{phot}$ for the non-irradiated disk (triangles). The disk and stellar parameters are the same than Figure \ref{fig_taus}. 
 }
\label{fig_temp}
\end{figure}

 Figure \ref{fig_temp} also shows that 
 the central temperature distribution becomes flat around 
$T_c \sim 1800 - 2000 \ \K$ 
 ($R \lesssim 0.1 \ \AU$)
where the Rosseland mean optical depth has a minimum (see Figure 
\ref{fig_taus}).
 The dust is the dominant opacity source at temperatures lower 
than the sublimation temperature, which depends on the 
type of grain  and the local  
 density. 
In the regions where dust is destroyed the Rosseland mean 
 opacity  $\chi_R$ 
 decreases.  At a given annulus,  
 the optical depth increases towards the 
 midplane (i.e., with a decreasing $z$), until a 
 depth is reached 
 where the dust is destroyed. 
From this height to the 
 midplane 
 the optical depth remains constant.
 In the region where dust is destroyed, 
 the temperature is almost constant with 
 height, being  $T_c \sim T_{sub}$ (see \cite{D96}).
 If the density were high enough to increase the optical depth 
towards the midplane (if there is a heating source 
at the disk midplane, e.g., viscous dissipation) then $T_c > T_{sub}$.
This is the case  of 
 disks with a lower $\alpha$ or a higher $\Mdot$ than the model presented in this paper.

As can be seen in Figure \ref{fig_taus}, the Rosseland  optical depth 
decreases with radius, so the gradient of temperature 
required to transport the energy produced in the disk also decreases 
with radius. The central temperature, for the optically thick annuli 
where the diffusion approximation holds, scales as   $T_c \approx 
T_{phot} \ (3 \tau_R/4)^{1/4}$.  
 For $R \gtrsim 10 \ \AU$,  where $\tau_R \lesssim 10$, 
the photospheric and the central temperatures becomes very similar,  
 i.e.,  
the disk {\it interior} is almost vertically isothermal. 
In this outer region, the temperature is  
given approximately by 
$T \approx 0.84 (F_{irr}/\sigma_R)^{1/4}$, as follows 
 from equation (\ref{eq1}), with $J=J_\infty=const$,  
 $T=const$,  assuming $\tau_P >> 1$, $F_{irr} >> F_{vis}$, $F_{ion}$,  
and neglecting the diffuse 
radiation field.

\begin{figure}
\plotone{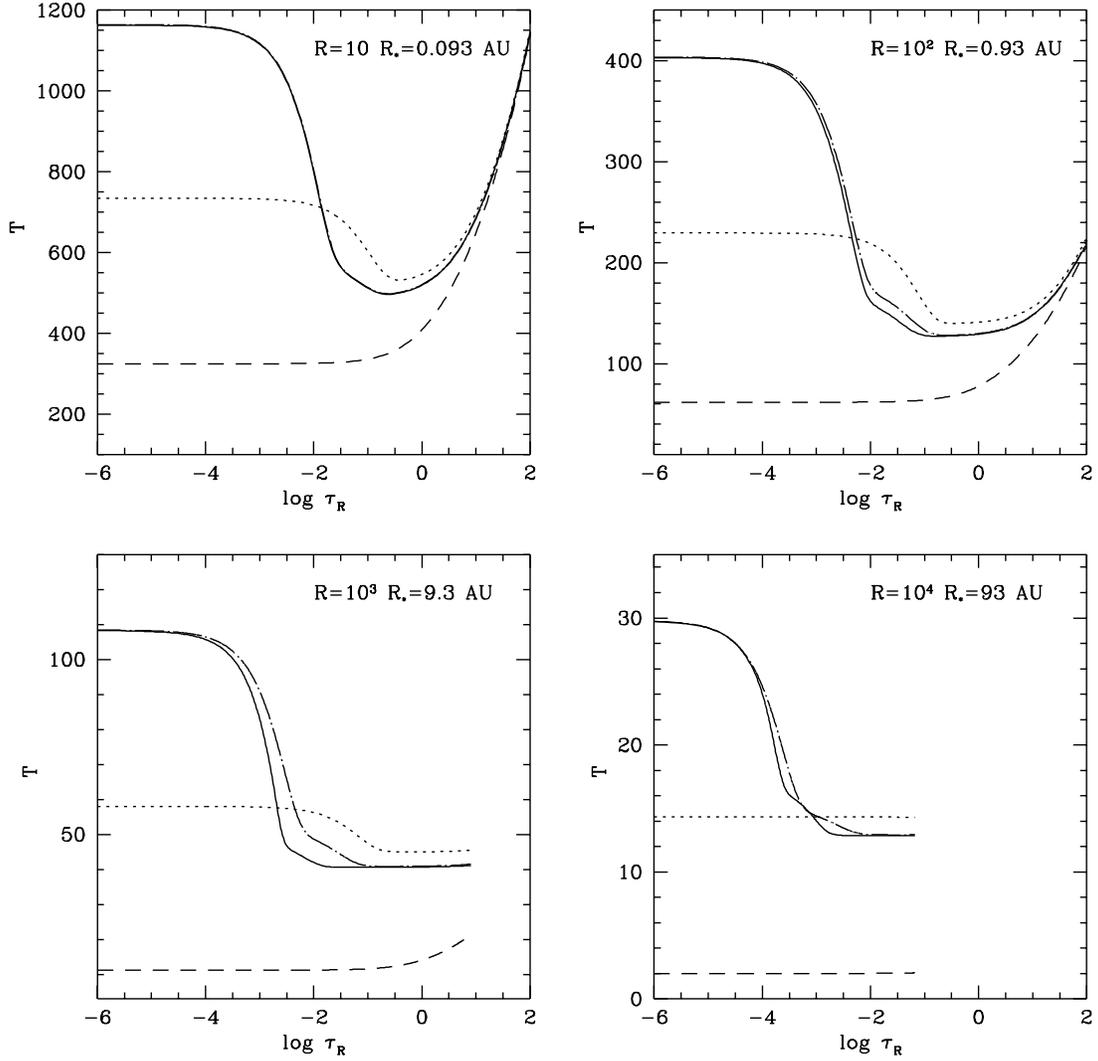}
\caption { Vertical temperature distribution of an irradiated disk 
 vs logarithm of the 
Rosseland mean optical depth (integrated from the disk surface to $z$). 
The curves represent: the temperature calculated in this paper (solid line), 
the temperature of a non-irradiated viscous disk (dashed line), 
the temperature given by CPMD (dot-dashed line), 
 temperature calculated
with the method of Malbet and Bertout (1991) (dotted line). 
 Irradiation produces a temperature inversion towards 
the
surface at all radii, the photosphere is at $\tau_R =2/3$.
The disk and stellar parameters are the same than Figure \ref{fig_taus}. 
}
\label{fig_vertical}
\end{figure}

At a given annulus, the density decreases with height and 
thus the optically thick annuli have 
upper optically thin layers, i.e.,  
 a {\it disk atmosphere}.
Figure \ref{fig_vertical} shows the vertical distribution of temperature of the 
reference model 
at different radii.  For all the disk there 
is a 
 {\it temperature inversion}  produced 
by the stellar irradiation (see CPMD). 
 This temperature inversion is due to 
 two effects, previously explored by CPMD:
(i) the stellar radiation penetrates the disk along a slanted direction 
so the optical depth along its trajectory is larger than the case of 
a penetration perpendicular to the disk surface (see also 
\cite{MB91}), and  (ii) the opacity 
 of the disk material (dominated 
 by dust) to 
the stellar radiation, 
  is larger than the disk 
opacity to its own radiation (due to the shorter frequencies 
of the former). Malbet \& Bertout (1991) 
considered only the first effect, but  the second effect 
is dominant as can be seen in Figure \ref{fig_vertical}.

The main difference between CPMD's treatment and the model presented in this paper is that we 
 calculate the disk optical depth self-consistently with the temperature and density vertical structure, while 
CPMD assume the disk interior is optically thick and that there is no 
viscous dissipation at the disk atmosphere, where energy is only transported.
 With the same set of equations, we calculate the optically thin atmosphere and the optically thick interior structures,
without any assumption {\it a priori} about the optical depth 
of the different regions. 
Annuli situated at large distances from the star, can be completely 
optically thin and the assumptions made by 
 CPMD are not valid, while with the method described in this paper, 
the structure of the optically thin outer disk can be calculated. 
Thus, we 
have found  that 
these regions have a temperature inversion, being  
the surface temperature  higher than the photospheric temperature. 
The annuli with low Rosseland mean optical depth, are almost 
vertically isothermal between the midplane and the photosphere, but 
 has a large contrast in temperature between the photosphere and the disk surface.

Figure \ref{fig_vertical} also shows that 
the approximation found by CPMD, with 
a correction introduced by the use of the Planck mean opacity 
in equation (\ref{eq_dfddz}),  
  is a good description of the temperature 
in the disk atmosphere.
 The disk upper atmosphere, with a temperature higher than the photospheric temperature, is like the ``super-heated'' layer proposed by 
Chiang \& Goldreich (1997), where the stellar radiation penetrating the disk is deposited.  
The main difference with their treatment and this paper 
is that here the disk vertical structure is calculated  in detail. 

\begin{figure}
\plotone{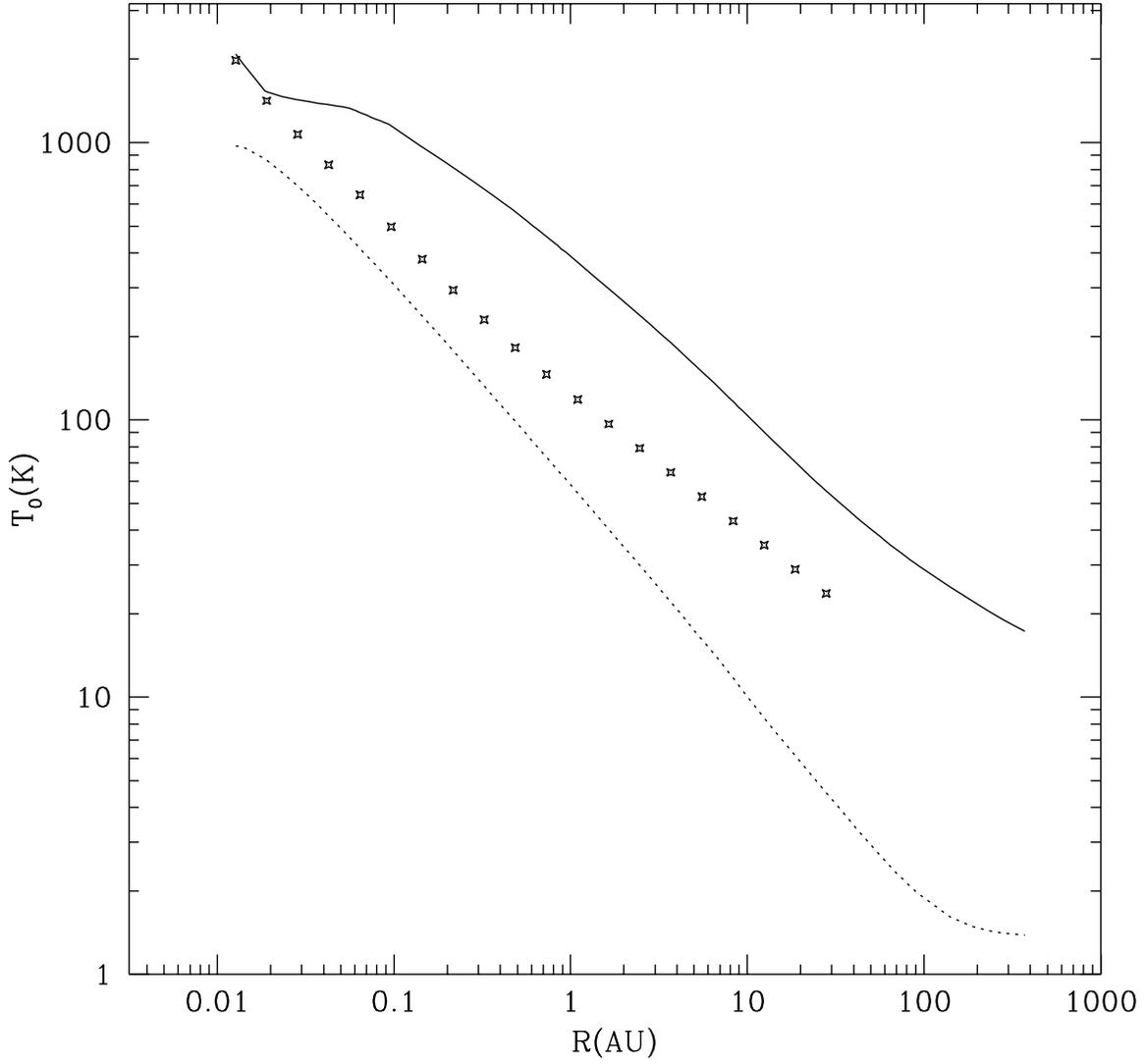}
\caption { Temperature at the disk surface $T_0$, for an irradiated disk 
(solid line) and a non irradiated disk (dotted line). The  
photospheric temperature $T_{phot}$ of the irradiated disk 
is shown as a reference (stars). The disk and stellar parameters are the same than Figure \ref{fig_taus}. 
}
\label{fig_tsup}
\end{figure}

Figure \ref{fig_tsup} shows the surface temperature of the disk $T_0$, 
i.e., the 
temperature at $z_\infty$.
 From $R \sim 0.5 \ \AU$ to $80 \ \AU$, $T_0$ is given approximately by
  $T_0 \sim R^{-0.56}$, and for larger radii  
we found 
 $T_0 \sim R^{-2/5}$, which is consistent with the expected 
temperature of optically thin dust 
(with an absorption coefficient $\kappa_\lambda \propto 1/\lambda$) heated 
by stellar radiation (\cite{spitzer},  pag. 194; see also 
\cite{CG97}).

\subsection {Characteristic Heights}

\begin{figure}
\plotone{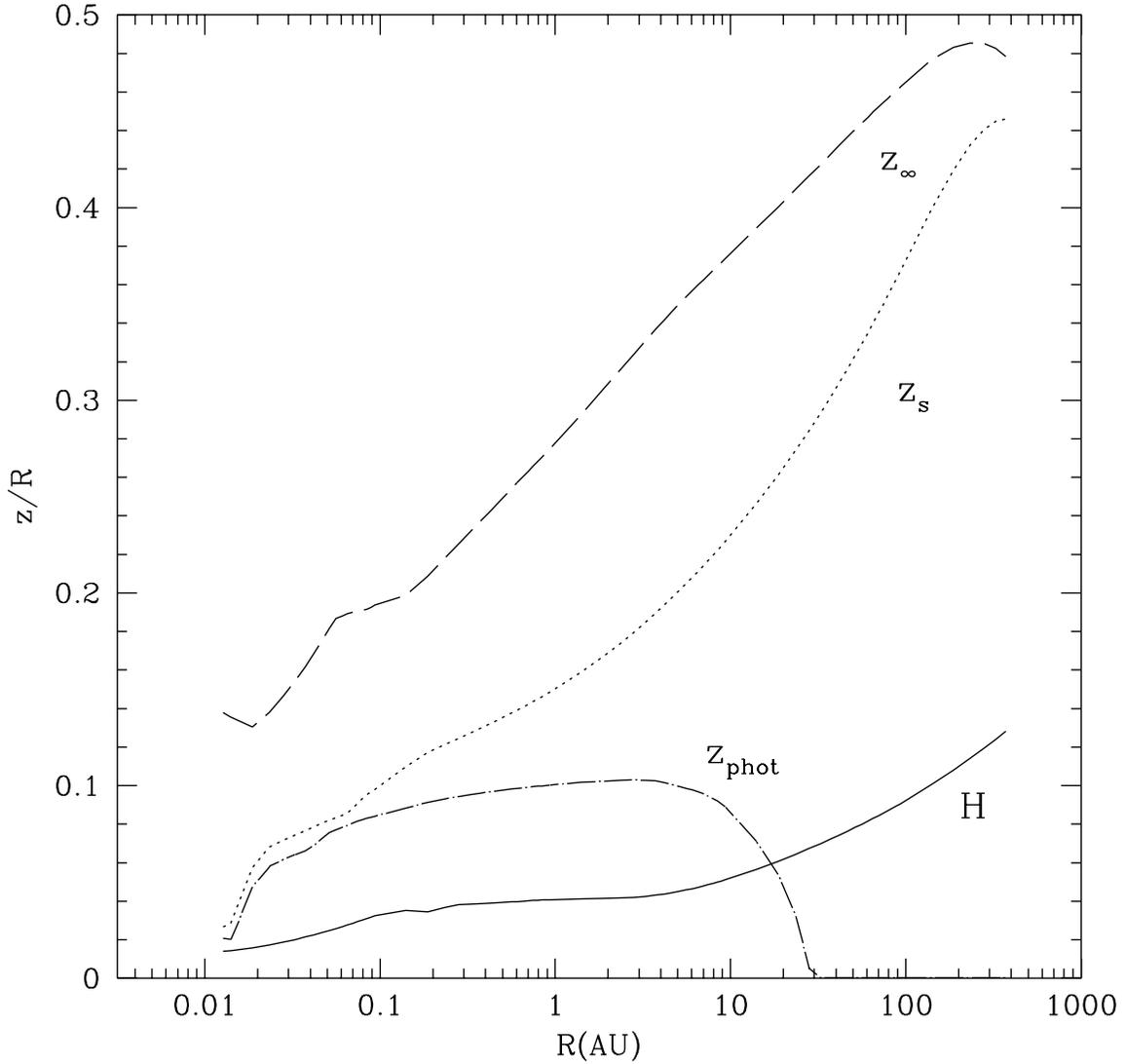}
\caption {Radial distribution of different 
 characteristic
heights of an irradiated 
accretion disk. 
The plotted heights are: the gas scale height $H$ evaluated at the central temperature (solid line), 
the disk surface height $z_\infty$ (dashed line), 
the height where a large fraction of stellar radiation is absorbed  $z_{s}$ 
(dotted line), and the photospheric height $z_{phot}$ (dot-dashed line), all
in units of the disk radius. The disk and stellar parameters are the same than Figure \ref{fig_taus}.  }
\label{fig_altura}
\end{figure}

There is not a unique definition of disk height.
Figure \ref{fig_altura} shows  the disk photospheric height $z_{phot}$, the 
scale height 
of the gas at the midplane temperature  $H$, 
the height at which the stellar radiation 
is deposited $z_s$, 
and the height at which $P=P_\infty$ given by $z_\infty$. There are clear 
differences between these heights which can be understood as follows.
 
The pressure scale height of the gas 
 at the central temperature
is given by: 

\be
{H \over R} = {c_s(T_c) \over R \Omega(R)} = \biggl [ {k T_c R \over  G M_* 
\mu(T_c,\rho_c) m_H } \biggr ]^{1/2}. 
\label{eq_scaleh}
\en

If $T_c/\mu$ decreases with $R$ slower than 
$1/R$,  
 then
the gas pressure scale height increases with $R$. 
 The higher the central temperature, the  larger the 
 pressure scale height. Therefore  stellar irradiation increases 
this scale  height with respect to a purely viscous disk.
For $R \gtrsim 6 \ \AU$, $H \propto R^{5/4}$ as it is expected from 
equation (\ref{eq_scaleh}) 
when $T_c \propto R^{-0.5}$. 

The photospheric height $z_{phot}$ is defined as the height
 where $\tau_R=2/3$, if the disk is optically thick.
 For $R \gtrsim 30 \ \AU$ the disk becomes optically thin and there is no 
photosphere.
The ratio $z_{phot}/H \approx 2.5$ between $0.1$ and $20 \ \AU$.

The absorption height  $z_{s}$, given by the height 
at which $\tau_s/\mu_0 = 1$, is the height where
 the largest fraction of stellar 
radiation is absorbed. 
 Figure \ref{fig_altura} shows that $z_s$ is larger than $z_{phot}$ at every 
radius 
of the disk, reflecting the fact  
that  the  disk becomes optically thick to the stellar radiation 
above the depth at which it becomes optically thick to its own radiation. 
For $R \gtrsim 20 \ \AU$ the disk is optically thick to the stellar radiation and optically thin to its own radiation.
 At large distances from the star, $z_{s}$ 
is close to the maximum height of the disk model $z_\infty$.  

From the vertical structure of this reference model, we find that  $z_s/H$ 
decreases 
from 4.5 to 2.5, when $R$ changes from $10$ to $300 \ \AU$.
The ratio
$z_s/H$ is assumed constant in models where there is no vertical structure 
calculation (e.g. KH, 
Cant\'o \etal 1995, \cite{CG97}) leading to   
$T_{phot} \propto R^{-3/7}$  instead 
of $T_{phot} \propto R^{-1/2}$.

The physical model considered in this paper is 
different  from that described by \cite{bell97}, where 
the disk is illuminated by a uniform radiation field with different 
temperatures ($T_R=10, 20 \rm{and} 100 \ \K$). 
In that case, the relevant height is the {\it vertical} optical depth at the 
characteristic wavelength range of the incident radiation. 
Bell \etal find for the lowest radiation temperatures, the height
decreases at large radii so that the outer disk regions cannot see the
star.
In  the model 
studied in this paper, he radiation is coming from the central star ($T_*=4000 \ 
\K$), with a large 
incident angle respect to the normal to the disk surface and also at shorter 
wavelengths. Both properties combine to produce a flared disk, with an 
absorption height increasing with radius. Therefore,  when stellar 
irradiation is included in a self-consistent way, taking into account its 
characteristic
wavelength and direction, the outer disk can be heated by the central star.

We can see that 
 all the 
 characteristic heights plotted in Figure \ref{fig_altura}
are less than $R$, supporting 
the thin disk approximation, used to decouple the vertical and radial 
structures.

\subsection {Surface density}

\begin{figure}
\plotone{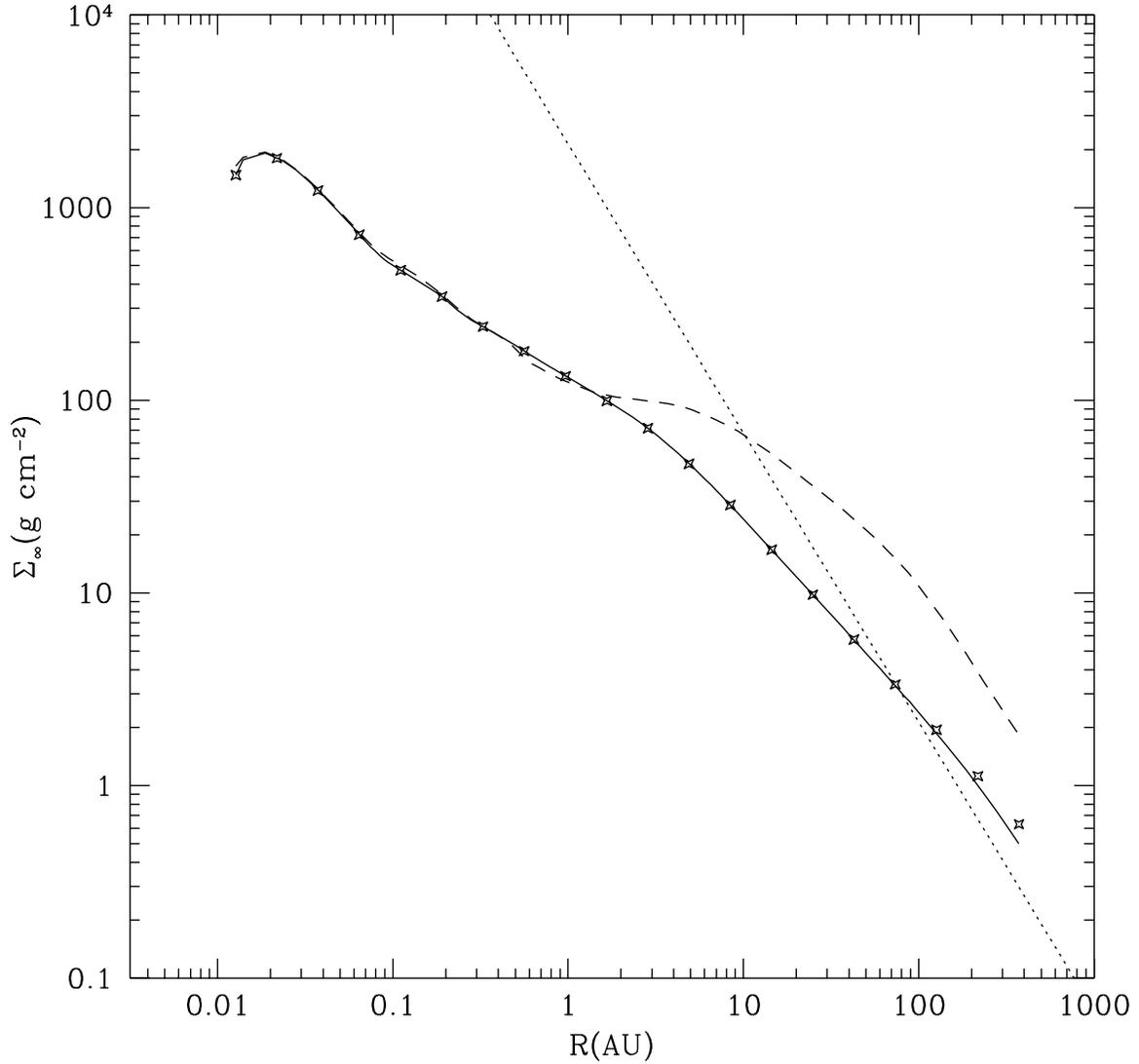}
\caption {Radial distribution of surface density of an
 irradiated (solid line)  
and a non irradiated (dashed line)  
disk model. For the irradiated disk, the approximation 
for  $\Sigma_\infty$ given by 
equation (\ref{eq_sigaprox}) is also plotted (stars). The nominal surface density $\Sigma \propto R^{-3/2}$, corresponding to the same total mass of the irradiated disk (assuming $R_d=373 \ \AU$), is also shown (dotted line) (see text). The disk and stellar parameters are the same than Figure \ref{fig_taus}. }
\label{fig_sigma}
\end{figure}

Given the disk density as a function of $R$ and $z$, 
the disk surface density radial distribution  
$\Sigma_\infty(R)$ is 
calculated using equation (\ref{eq_sigma}).
 Figure \ref{fig_sigma} shows $\Sigma_\infty(R)$ for  
an irradiated and a non irradiated disk 
with the same $\Mdot$, $\alpha$ and central star. 
From the equation of conservation of angular momentum in a 
steady $\alpha-$disk, the surface density
can be written as:

\be
\Sigma_\infty = {\Mdot \over 3 \pi <\nu_t>} \biggl [ 1 - \biggl ( {R_* \over R} \biggr )^{1/2} \biggr ]
\en 

\noindent
where $<\nu_t>= \int_{-\infty}^{\infty} \nu_t(z,R) \  \rho(z,R) dz/ 
\Sigma_\infty$ is a mean viscosity coefficient.
 Approximating $<\nu_t>$  by the viscosity coefficient evaluated at 
the disk midplane, 
 the surface density can be written as:

\be
\Sigma_\infty \approx {\Mdot \over 3 \pi \alpha }{\Omega \mu m_H \over k T_c} 
[ 1 - \biggl ( {R_* \over R} \biggr )^{1/2} \biggr ]
\label{eq_sigaprox}
\en

Figure \ref{fig_sigma} shows this expression for 
$\Sigma_\infty$  
compared with 
the surface density calculated numerically from $\rho(R,z)$.
The approximation is a good description of 
the numerical result, which can be used as a check of the method 
used to integrate the vertical structure equations. 
From $R=2$ to $100 \ \AU$, $\Sigma_\infty \sim R^{-1}$, as it is expected 
from 
equation (\ref{eq_sigaprox}) when $T_c \sim R^{-0.5}$.
Then, a steady $\alpha$-disk has a surface density distribution flatter than the nominal form $\Sigma_\infty= \Sigma_0 \ (R/R_*)^{-3/2}$, usually found in the literature (e.g. \cite{BSCG90}; \cite{dutrey96}; \cite{MN95}, 
\cite{CG97}).
 Figure \ref{fig_sigma} also shows 
this power law surface density for a disk with the same 
total mass (assuming $R_d=100 \ \AU$)  than our reference model. 
The $\alpha$ irradiated disk has  denser outer regions than the nominal model. 
This implies that the brightness of the outer disk is higher (for a fixed 
temperature distribution and opacity) at submm and mm wavelengths, where  
the disk becomes optically thin.  This result has implications on 
the physical properties of disks as inferred from observations at long wavelengths, 
which will be discussed in paper III.

\subsection{Disk mass}

The mass of a disk model is given by:

\be
M_d = \int_{R_i}^{R_d} \Sigma_\infty 2 \pi R dR 
\label{eq_masa}
\en

\noindent
where $R_i$ is the inner radius of the disk, which can be $R_*$
 if the disk extends inward to the surface of the star, or 
the radius of an inner hole $R_{h}$, when a magnetosphere is present (e.g. \cite{GL79}), 
and 
   $R_d$ is the 
outer radius of the disk.
 Due to the increase in surface area, 
$M_d$ is determined by the mass of the outer regions, 
in spite of the fact that the surface density decreases with $R$, as discussed 
above.
From equation (\ref{eq_sigaprox}) we can see that unless the unlikely case 
of the central 
temperature increasing with $R$, 
the area of the annuli increases with $R$ more rapidly than the decrease 
in $\Sigma_\infty$. This result implies that the disk mass is not sensitive 
to the value of $R_i$ (if $R_i << R_d$) and 
that the heating mechanisms other than viscosity, 
 affect strongly 
the mass of a disk model with fixed values of $\Mdot$ and
$\alpha$.

\begin{figure}
\plotone{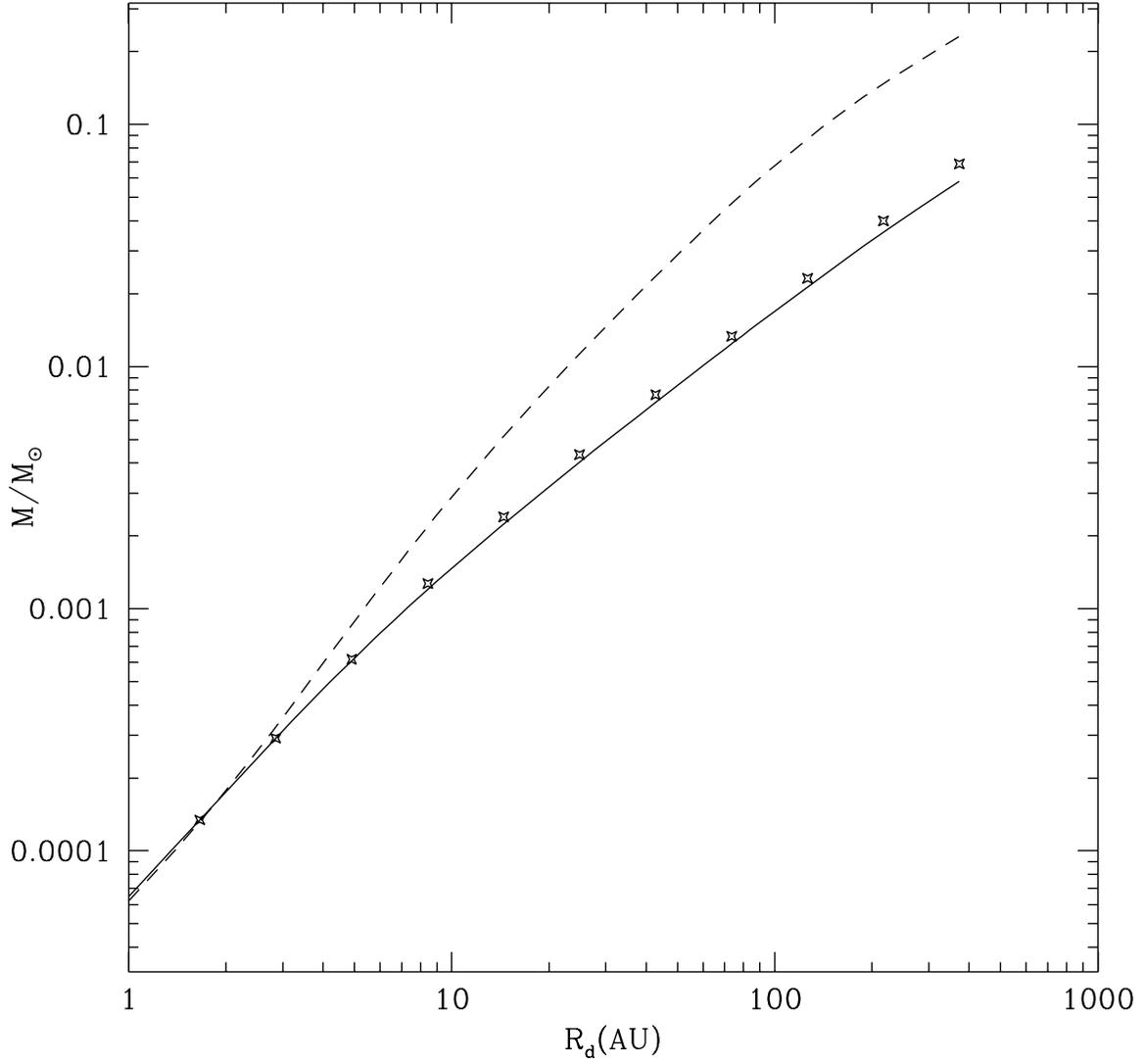}
\caption {Radial distribution of total mass of an irradiated (solid line) 
and a non irradiated (dashed line) 
disk model. 
The mass of the central star is $M_*=0.5 \ \MSUN$. For the irradiated disk,
 the approximation given  by equation (\ref{eq_masaprox}) is also plotted 
(stars). The disk and stellar parameters are the same than Figure \ref{fig_taus}. }
\label{fig_masa}
\end{figure}

Figure \ref{fig_masa} shows the mass of  
the  disk as a function of $R_d$. At $R_d=100 \ \AU$ 
the disk mass is $M_d \approx 0.017 \ \MSUN$, 
 larger than the Minimum Mass of the Solar nebula, $\sim 0.01 \ \MSUN$, but 
smaller than the central star mass $M_* = 0.5 \ \MSUN$. 
Therefore, it seems to be a good approximation to neglect 
the disk 
  self gravity 
with respect to the stellar gravity, as was assumed when 
the angular velocity of the disk  was taken as the 
 Keplerian 
angular velocity and, also, in the hydrostatic equilibrium equation.

Based on the result that the disk mass is determined by its outer regions, 
a rough estimation of $M_d$ 
 can be constructed assuming that the disk is vertically isothermal.
Taking the midplane temperature as  a power law,  
$T_c \approx 
T_1 (R/R_1)^{-\gamma}$ for $ R \ge R_1$ 
with
$T_1=T_c(R_1)$,  and equations (\ref{eq_sigaprox}) and (\ref{eq_masa}),
the disk mass can be written as:

\be
M_d \approx M_{inner}(R_1) + { 2\mu m_H \over 3 k T_1} 
\sqrt{G M_* R_1 } {\Mdot \over \alpha} (\gamma +1/2)^{-1} 
\biggl [ \biggl ( {R_d \over R_1} \biggr )^{\gamma+1/2} -1 \biggr ]     
\label{eq_masaprox}
\en

\noindent
where $M_{inner}(R_1)$ is the disk mass corresponding to regions with $R < R_1$, 
which we find is $M_{inner}(R_1) << M_d(R_d)$ if $R_d >> R_1$. 
For the reference model 
we find $R_1 \approx 456 \ R_*$,  $T_1 \approx 58 \ \K$ and $\gamma \approx 
0.5$.
 The 
mass $M_{inner}(R_1)=5 \times 10^{-4} \ \MSUN$, and  $M_d \propto R_d$.
Figure \ref{fig_masa} shows the approximation described by equation 
(\ref{eq_masaprox}) 
compared with the numerical result.

\subsection {Characteristic timescales}

The timelife of 
a disk with  $\Mdot=10^{-8} \ \MSUNYR$ and  $M_d = 0.017 \ \MSUN$ 
is $M_d/\Mdot \approx 1.7 \times 10^{6} \ \yrs$.
The characteristic time of  the matter diffusion due to 
viscous torques $t_{vis}$, is:

\be
t_{vis} \approx {R \over |u_R|} \approx {R^2 \over <\nu_t>} = {R^2 \Omega \over 
\alpha c_s(T_c)}
\label{eq_tvis}
\en

\noindent
where $u_R$ is the radial accretion speed. 

\begin{figure}
\plotone{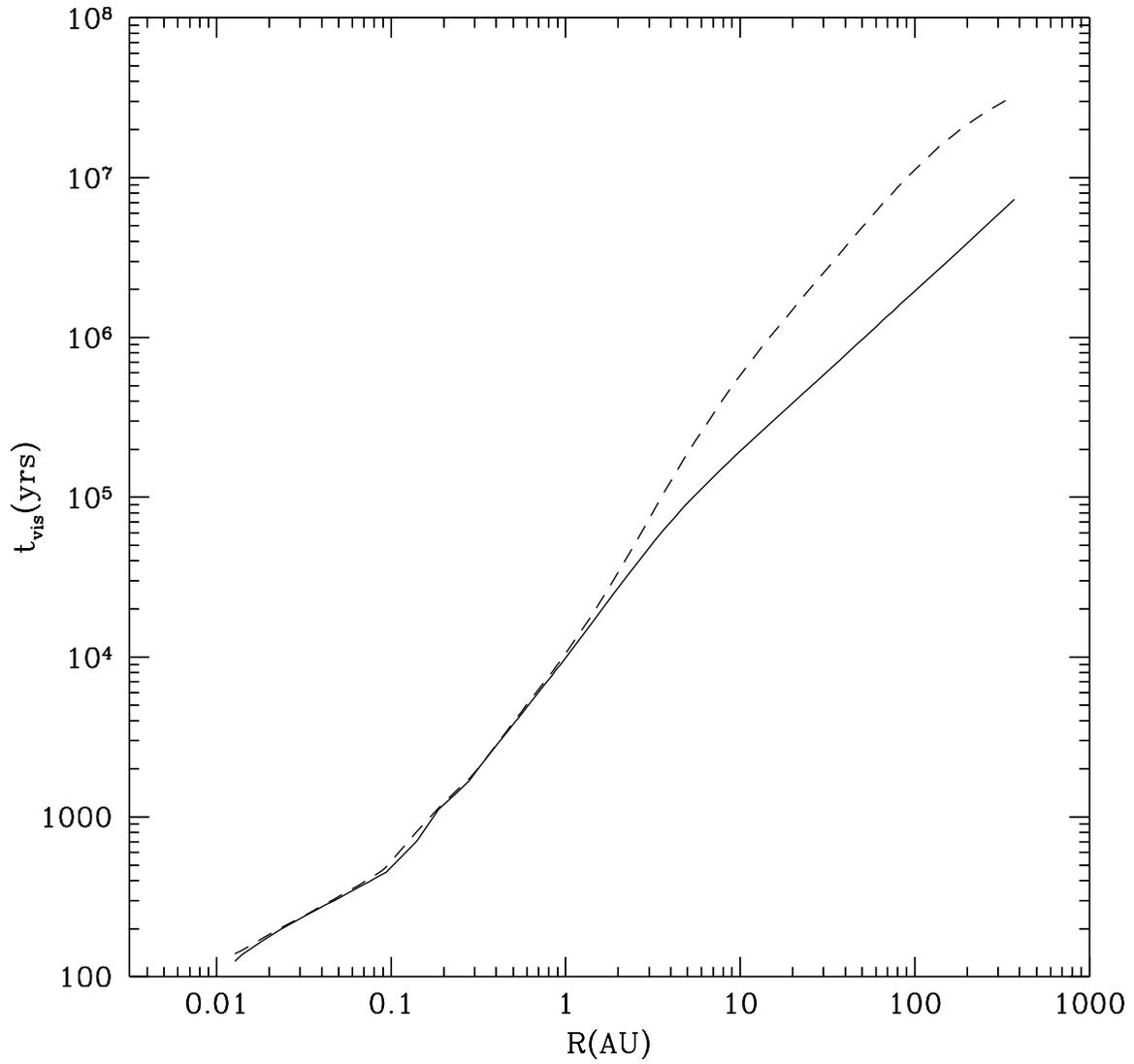}
\caption {Radial distribution of the viscous time for an
 irradiated (solid line) and a non-irradiated (dashed line) disk as a function 
of radial distance to the central star.  The mean disk age 
estimated (Strom \etal 1995) is $\sim 10^{6} \ \yrs$. The disk and stellar parameters are the same than Figure \ref{fig_taus}. }
\label{fig_tvis}
\end{figure}

The viscous time evaluated at $R_d$ is an estimation of the lifetime of a disk 
which receives 
little mass from the surrounding cloud. Figure \ref{fig_tvis} 
shows that the reference model 
at $100 \ \ \AU$
has  $t_{vis} \approx 2 \times 10^{6} \ \yrs$, also consistent with 
the 
 disk 
mean age $\sim 10^{6} \ \yrs$, inferred from the position of T Tauri stars  
in the HR diagram (Strom \etal 1995). 

From equation (\ref{eq_tvis}) we can see
that the larger the central temperature the shorter the lifetime of the disk,
 for a given value of $\alpha$. 
Therefore, the stellar irradiation, which is the most important 
 outer disk heating mechanism, 
    can not be neglected 
in a lifetime estimation of a given disk model.

\subsection {Energy transport}

\begin{figure}
\plotone{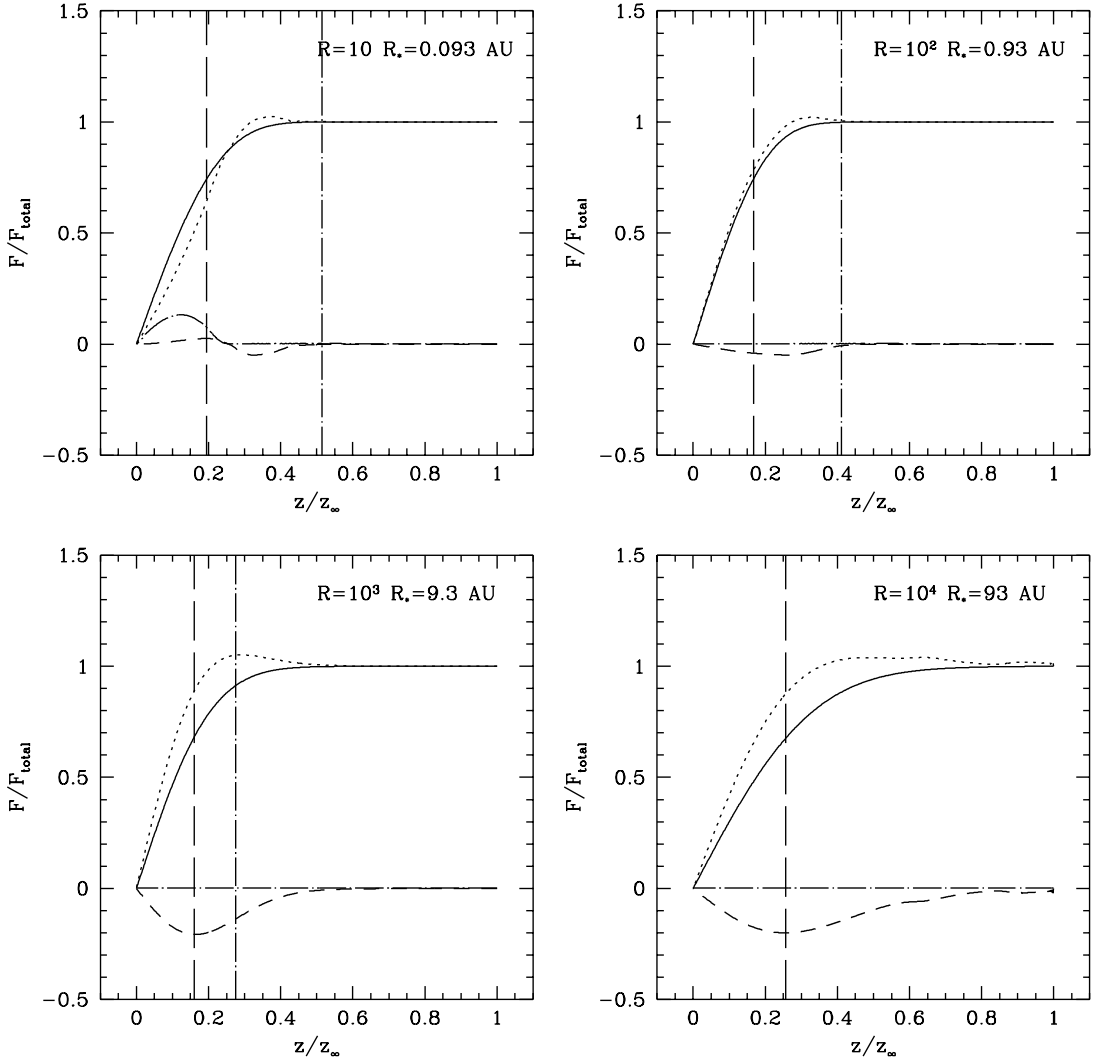}
\caption {Vertical distribution of the fraction of total energy 
flux transported by  different mechanisms, for different annuli. 
The fluxes are: 
total intrinsic flux (solid line), 
radiative flux (dotted line), turbulent flux 
(dashed line) and convective flux (dot-dashed line).
The position of the gas scale height 
evaluated at the midplane temperature  $H$ 
is shown with a vertical dashed line and the 
position of the photosphere is shown with a vertical dot-dashed line. 
Each panel shows at the upper corner the radius 
of the given annulus
 in units of stellar radius.  The disk and stellar parameters are the same than Figure \ref{fig_taus}.  }
\label{fig_fluxes}
\end{figure}

Figure \ref{fig_fluxes}  shows 
 the fraction of the total energy flux 
transported by radiation, convection and turbulence, as a function 
of height, 
 at various distances 
from the central star. 
 Also the position of the gas pressure
 scale height $H$ is shown.

A flux with a positive sign is going in the $+z-$direction, and with 
a negative sign, is going in the $-z-$direction.
 In the convective regions (see the panel corresponding to $R=0.093 \ \AU$ of Figure 
\ref{fig_fluxes}) the turbulent flux is positive, thus
  it is transporting energy from inside to outside (form inner regions to upper regions).

Convection only transports an appreciable amount of
 energy near the midplane around  
 $R =  10 \ R_* \sim 0.01 \ \AU$. 
In the radiative 
zones, the turbulent flux becomes negative, as  can be seen 
from equation (\ref{eq_fturb}) when $\nabla < \nabla_A$. 
In these  radiative or  non-convective regions, 
the turbulent flux transports energy from the upper layers, where 
temperature and pressure are low, to the inner layers, which are 
hotter and have a higher pressure.
When a turbulent adiabatic 
eddie moves in a subadiabatic medium  from a cold to a hot region, it will 
have a higher temperature respect to 
the surrounding medium. On the other hand, if 
this adiabatic eddie moves from a hot to a cold region, it will 
have a lower temperature respect to the surrounding medium. 
The work made by the medium on the element or by it on the medium, is responsible for this behavior. 
Therefore, the presence of the turbulent flux increases the radiative flux, as can be seen in Figure \ref{fig_fluxes}.  
At every annuli, we can see that radiation is the 
most important energy transport mechanism inside the disk, 
enhanced in the non-convective regions  because the presence of the turbulent flux. 
Thus, radiation is 
controlling the disk temperature structure.   

Note that less than 70 \%  
of the total flux is produced below the gas scale height. On the other hand,  
 above the disk photosphere, the total flux is almost constant (see the 
panels corresponding to $R=0.093, \ 0.93, \ 9.3 \ \AU$ 
from Figure \ref{fig_fluxes}). So,   
 it seems to be a good approximation to neglect the flux produced there 
by viscous dissipation and ionization due to energetic particles. Then,  
in the 
disk atmosphere,  the energy is being only transported.

\subsection {Disk gravitational stability} 

Other interesting physical property that can be calculated from a detailed 
disk model is the gravitational stability parameter introduced by Toomre (1964), 
which  quantifies the  fluid reaction against axisymmetric gravitational
 perturbations.
 The Toomre parameter can be written as:

\be
Q_T = {c_s \Omega \over 2 \pi G \Sigma_\infty}, 
\en

\noindent
such that, if $Q_T > 1$ the fluid is stable.

\begin{figure}
\plotone{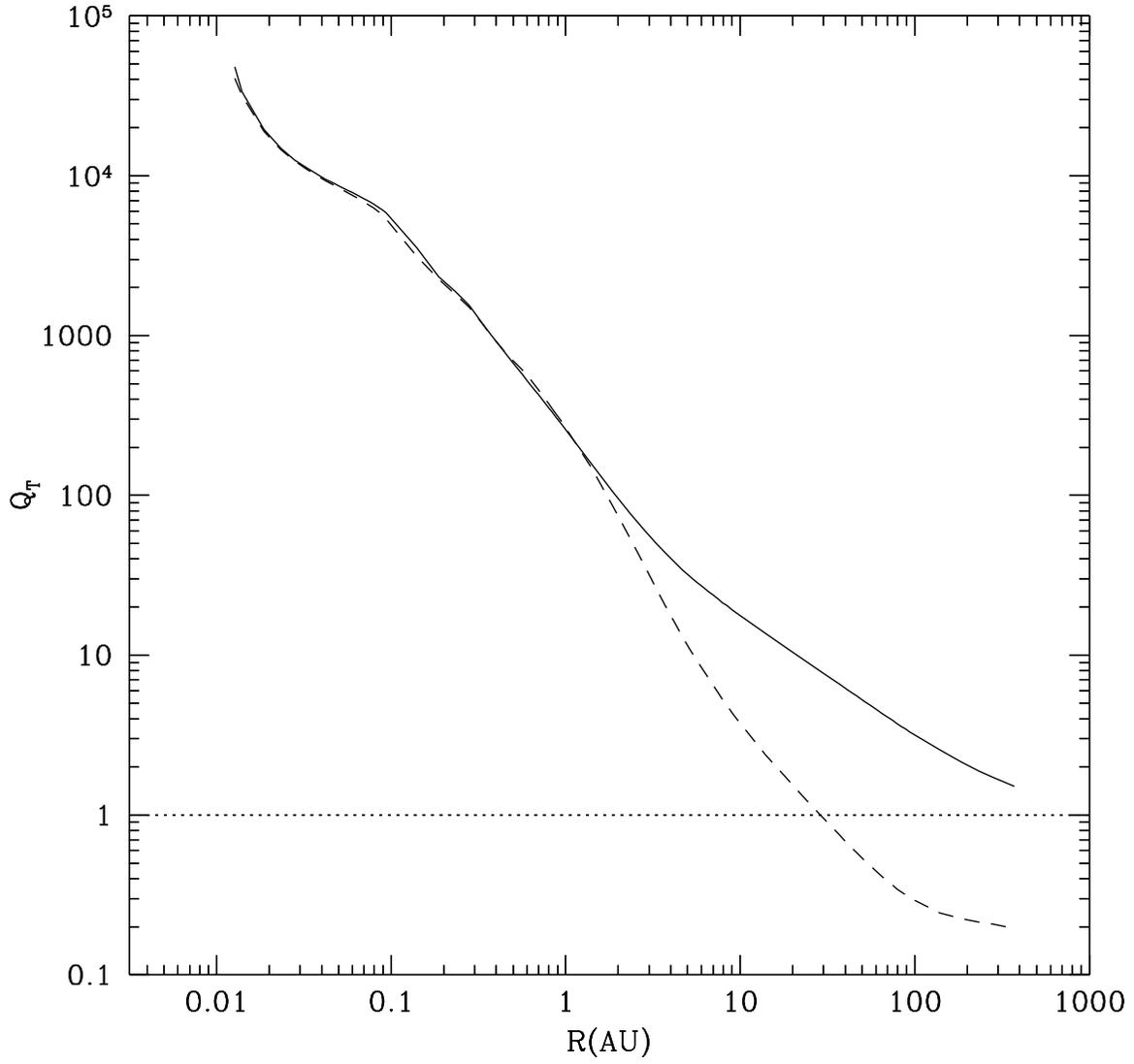}
\caption { Radial distribution of Toomre's instability parameter $Q_T$
 for an irradiated (solid line) and a non-irradiated (dashed line) disk model. 
The value $Q_T=1$ corresponds to the limit between stable and unstable regions 
(dotted line). The disk and stellar parameters are the same than Figure \ref{fig_taus}.  }
\label{fig_toomre}
\end{figure}

Figure \ref{fig_toomre} shows $Q_T$ as a function of $R$ for the reference 
model;
the disk is stable for the range of radii used to calculate 
the disk structure ($R \le  337 \ \AU$). It 
is interesting to note that if the irradiation flux  is 
higher  $Q_T$ increases due to  the increase in 
 disk temperature and the decrease in  surface density. 
 Thus, irradiation tends to stabilize the disk against 
gravitational perturbations.

\subsection{Ionization structure}

Given the vertical structure, it is interesting to evaluate where  
the ionization fraction (controlled by energetic particles and 
thermal ionization) 
is not enough to sustain the BH instability (\cite{BH91}). Gammie 
(1996) has called 
this region 
``the  disk dead zone'',  where 
turbulent viscous dissipation cannot exist if the source of turbulence is 
the mentioned magnetic instability.

Following Gammie (1996), the ionization fraction 
which corresponds to a magnetic Reynolds number $Re_{M}=V_A H / \eta=1$, is 
given by:

\be
x_1=0.135 \times 10^{-13} \ \alpha^{-1/2} \ \biggl ( {R \over 1 \ \AU} \biggr 
)^{-3/2} \ \biggl ( {T \over 500 \ K} \biggr)^{-1} \ \biggl ( {M_* \over \MSUN} 
\biggr)^{1/2}
\en

\noindent
where $V_A$ is the Alfven speed, approximated as $V_A \approx \alpha^{1/2} 
c_s(T)$, $\eta$ is the resistivity, given by $\eta=6.5 \times 10^{3} x^{-1} \ 
\cm^2 \ \seg^{-1}$, $x=n_e/n_H$ is the ionization fraction, and 
$n_e$ is the electronic density.

On the other hand, from Stepinski (1992), the local ionization fraction  due 
to 
 energetic particles, considering recombination upon grain surfaces and ion 
reactions,
 can be written as:

\be
x_{ep}=5.2 \times 10^{-18} \ r_{gr}^{-1} \ T^{1/2} \biggl ( 
\sqrt{1+10^{-7} {r_{gr}^2(\zeta_{rad}(^{26}Al)+\zeta_{cos}e^{-\Sigma/\lambda}) \over \rho^2 T}}-1 \biggl )
\label{eq_xep}
\en
where $r_{gr}$ is the grain typical size (we adopt $r_{gr}=0.1 \ \mu m$). 
 The cosmic rays ionization rate 
decreases exponentially with the surface density, with a 
characteristic attenuation density given by $\sim 100 \ \gr \cm^{-2}$.
 The thermal ionization fraction $x_{th}$ is calculated assuming LTE, and 
using the procedure described by Mihalas (1967). We approximate 
the ionization fraction in the disk as $x \approx max(x_{th},x_{ep})$.

\begin{figure}
\plotone{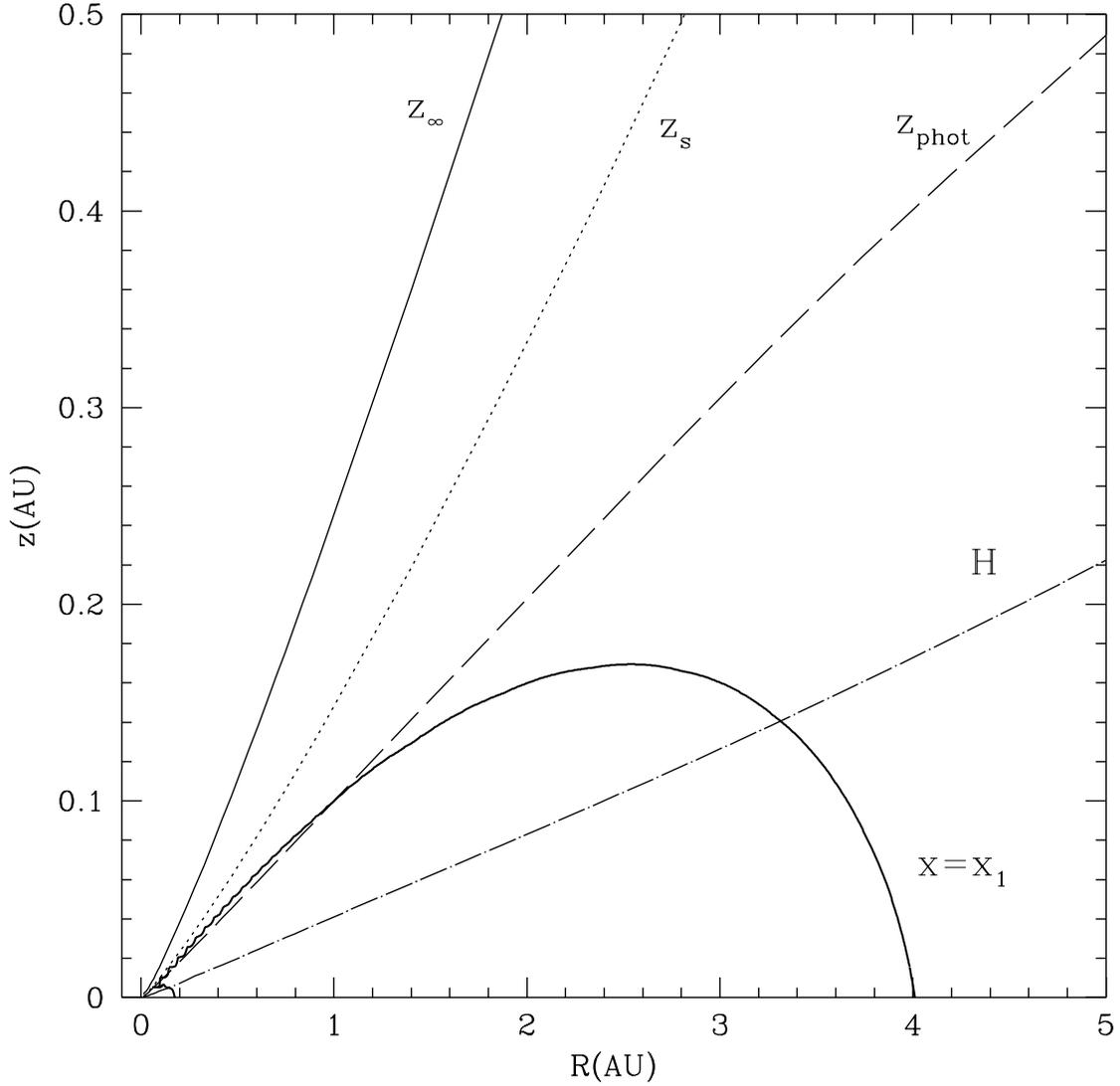}
\caption {The contour of $x/x_1=1$, where $x$ is the disk ionization 
fraction and $x_1$ is the ionization fraction corresponding to a magnetic Reynolds number $Re_{M}=1$. The region inside this contour, $0.2 < R < 4 \ \AU$ and 
$z < 0.17$, has $Re_{M} < 1$. 
We also plot the disk height $z_\infty$ (solid line), the height where a large fraction of the stellar radiation is absorbed $z_s$ (dotted line), the photospheric height $z_{phot}$ (dashed line) and the gas pressure scale height evaluated at the midplane $H$ (dot-dashed line)}
\label{fig_gammie}
\end{figure}

Figure \ref{fig_gammie} shows a contour of $x/x_1=1$.
In the region where $x < x_1$, the disk material is decoupled from the magnetic field, 
$Re_M<1$ and the BH instability cannot be the source of turbulence. 
This is the 
 ``dead zone'', where there can be no viscous dissipation, 
and mass transport. 
In spite of the disk heating due to stellar irradiation, the thermal ionization is not enough in this region to make $Re_M<1$.
 If there is no other source of turbulence, the region 
between $0.2$ and $4 \ \AU$ cannot be treated with a constant $\Mdot$ and 
$\alpha$. Outside this region, the magnetic Reynolds number is $Re_M>1$ and 
the disk turbulence can be initiated by the BH instability. 

As can be seen in Figure \ref{fig_gammie} the region where 
$x < x_1$ extends above the disk scale height, but it remains below the 
photosphere and the height where the stellar radiation is deposited. 
For $R \approx 2.5 \ \AU$, the surface density in the dead zone is 
$\approx 90 \ \%$ of the total surface density at this radius.
 At $R \sim 2 \ \AU$ the surface density drops below the attenuation scale for cosmic rays, $\sim 100 \ \gr \ \cm^{-2}$, 
but the ionization fraction implied by equation (\ref{eq_xep})
is smaller than the upper limit used by Gammie (1996), and $x_{ep}$ remains 
less than  $x_1$ 
between $2$ and $4 \ \AU$.

\subsection {Spectral energy distribution}

For the reference model we have calculated the SED, assuming the disk 
is pole-on. The maximum and minimum radii adopted to calculate the SED are $R_d=300 \ \AU$ and 
 $R_{hole}=1 \ R_*$, respectively.
For the millimeter-wave dust opacity, 
we use a power law given by:

\be \kappa_\nu = \kappa_0 \biggl ( {\lambda \over 200 \mu m} \biggr )^{-\beta},
\en
 
\noindent
where $\kappa_0=0.07 \ \cm^2 g^{-1} $ is obtained assuming that at wavelengths shorter
 than $200 \ \mu m$ the dust opacity is given
by Draine \& Lee (1984). The exponent $\beta$ is taken as 1.
 This opacity is, at $\lambda=1.3 \ mm$
 a factor of 2.14 smaller than the opacity used by Beckwith \etal (1990).

Figure \ref{fig_sed} shows the predicted SED compared with the SED of AA Tau, a 
typical Classical T Tauri star and the agreement is good for $\lambda > 2 \ \mu 
m$.  We want to emphasize that this is not a best fit model for AA Tau, is only 
a typical model. In paper III we will use observational restrictions as
the mass accretion rate and central star properties inferred from 
optical
 observations to restrict 
the family of models for different particular objects. 
Nevertheless, the comparison between the SED of a typical model with 
the SED of a typical CTTS show that the model predictions agree well with observations.

\section{Effect of stellar irradiation on the disk vertical structure}

Disks can be irradiated by external sources, such as  
 an infalling dense envelope (\cite{butner94}; \cite{DCH97}) 
  or 
a tenuous dusty envelope, remnant 
of the infalling envelope or associated with a disk or a stellar wind (\cite{natta93}),  
or the central star (\cite{AS86}; KH; 
CPMD and this paper). 
Different sources can be important in different evolutive phases 
of a disk, and in this paper we explore the case of 
a disk irradiated by the 
central star.
A non-irradiated model is
 used to quantify the effects 
of stellar irradiation  on the disk structure and physical properties,  
 although it does not 
represent a realistic disk for the case treated here (with $\Mdot=10^{-8} \ 
\MSUNYR$).

 Irradiation is the main heating source of the disk. 
Also  the heating due to ionization by energetic particles becomes 
larger than the viscous heating for $R > 200 \ \AU$, but it is negligible 
respect to the stellar irradiation. This behavior of the flux   is reflected  
in the photospheric temperature distribution, plotted in Figure \ref{fig_temp}.
For $R \lesssim 0.5 \ \AU$, $T_{phot}$
is similar to  that of an irradiated ``flat disk''. For larger distances, 
becomes flatter than the temperature distribution of a viscous 
disk or a flat irradiated disk, 
$T_{vis} \sim T_{flat}  \propto R^{-3/4}$.
 As shown in Figure \ref{fig_sed},  the irradiated disk has a 
larger 
emergent flux and a flatter SED  
 at every wavelength,  than 
the non-irradiated disk.
 For $R \lesssim  2 \ \AU$, $T_c$ is the same 
for the irradiated and non-irradiated models, because the viscous dissipation 
is the dominant  
 heating source of the regions close to the  disk midplane. 
At larger distances, the stellar radiation can penetrates the disk and it 
becomes the most important heating source at every height.

Figure \ref{fig_sigma} shows the surface density of the irradiated and 
non-irradiated disk model, with the same $\Mdot$ and $\alpha$. 
 Stellar irradiation 
decreases the disk surface density for $R \gtrsim 1 \ \AU$.
As can be seen 
from equation (\ref{eq_sigaprox}), the change in disk temperature 
caused by the irradiation substantially increases the viscosity, and so, for a 
fixed mass accretion rate, the surface density drops correspondingly. 
The mass of the disk, determined by its surface density 
at large radii, also decreases with irradiation. 
As can be seen in Figure \ref{fig_masa}, the mass of a pure viscous disk with 
$\Mdot=10^{-8} \ \MSUNYR$, $\alpha=0.01$ and
 $R_d=100 \ \AU$, is $M_d=0.063 \
 \MSUN$, approximately a factor of $4$ larger than an irradiated disk with 
the same $\Mdot$ and $\alpha$.
From equation (\ref{eq_masaprox}), a non irradiated disk 
has $M_d \sim R_d^{7/4}$, assuming $\gamma \approx 3/4$ in the outer regions, 
so its mass increases with the disk radius more rapidly than the mass 
of an irradiated 
disk   $M_d \propto R_d$.

The increase in temperature produced by irradiation
decreases the viscous 
time of the disk, as can be seen in Figure \ref{fig_tvis}. The non-irradiated 
disk has $t_{vis} \approx 10^{7} \ \yrs$, an order of magnitude larger than 
the estimated mean timelife of disks (\cite{SES93}).

Another property for which the effect of irradiation is 
important is the Toomre instability parameter. Figure \ref{fig_toomre} 
shows  that the non-irradiated disk becomes 
unstable for $R \gtrsim 28 \ \AU$.
The higher temperature and
lower surface density of an irradiated disk, with respect to a pure viscous 
disk with the same $\alpha$ and $\Mdot$, stabilizes the disk against 
gravitational perturbations for (at least)  
 $R \lesssim 337 \ \AU$.

\section{Effect of irradiation from an accretion ring on the disk structure}

The magnetospheric model has become  
widely accepted (e.g. \cite{konigl91}; \cite{CH92}; \cite{shu94}). 
 It is expected that the presence of a 
magnetosphere modifies the disk model discussed 
in this paper. Two important  
effects  on the disk structure and emission are:  
(1) to truncate the disk, such that the resulting inner 
hole decreases its  near IR emission and 
(2) to heat the disk, changing its temperature and
 surface density distributions.
The disk heating is produced by the the interaction with the 
 stellar magnetic field (\cite{KYH96}) and also, by radiation from an accretion shock at high stellar latitudes. 
 For low mass stars, this radiation is characterized by
photons of much shorter wavelength than those of the stellar radiation
or the local disk emission. Kenyon \etal (1996)  
conclude that the effect of this ring radiation 
is small, but they do not include the increased absorption 
efficiency due to the wavelength-dependence of the opacity.  
In this section, we study   the effect of the accretion ring radiation 
on the vertical structure of the reference model. 
The dependence of this additional heating source on disk and stellar 
parameters and its effect on 
physical (mass, stability, etc.) 
and observable (near IR colors, long wavelength 
fluxes, etc)  properties of disks will be calculated 
 in papers II and III.  

The calculation of the ring flux $F_{A}$   
intercepted by the disk and its mean incidence angle $\cos^{-1} \mu_A$
 is described in 
the Appendix. The ring is located at an angle $\theta_0$, measured from the $z$ axis (see Figure \ref{fig_annulus})  and has an angular width $\Delta \theta_0$. 
 We assume the ring has a temperature $T_A=10000 \ \K$, inferred from
 the UV excess of T Tauri stars (e.g. \cite{BBB88}). 
The ring angular width can be estimated as:

\be
\Delta \theta_0=0.057 \ (\sin \theta_0)^{-1} 
\biggl ( {L_A \over \LSUN} \biggr ) \biggl ({T_A \over 10^{4} \ \K} 
\biggr)^{-4} \biggl ({R_* \over 2 \ \RSUN} \biggr)^{-2}  \ \ rad
\en

\noindent
where $L_A$ is the ring luminosity (for a ring seen pole-on by the observer). 
  The ring luminosity can be related to 
the mass accretion  rate (see Kenyon \etal 1996), writing:

\be
L_A=0.019 \biggl ({M_* \over 0.5 \ \MSUN} \biggr)  
\biggl ({\Mdot \over 10^{-8} \ \MSUNYR}  \biggr) \biggl ({R_* \over 2 \RSUN}\biggr)^{-1} \biggl ( 2 - {R_* \over R_{hole}} \biggr ) \ \ \LSUN
\label{eq_la}
\en

The disk adopted as a reference model, has $\Delta \theta_0=0.062^\circ / (\sin \theta_0)^{-1} \ (2-R_*/R_{hole})$. We have taken two values for $\theta_0=30^\circ,60^\circ$, and an extreme value $(2-R_*/R_{hole})= 2$, to explore 
the effect of the irradiation of a  ring on the disk structure. The corresponding ring widths are: $\Delta \theta_0=0.25^\circ$, for $\theta_0=30^\circ$, 
and $\Delta \theta_0=0.14^\circ$, for $\theta_0=60^\circ$.
It is clear from equation (\ref{eq_la}) that the higher the $\Mdot$, the larger the ring luminosity, and the larger effect on the disk, at least 
on its upper layers or outer regions, where viscous dissipation can be less important.

We preserve the same set of equations and boundary conditions given in \S 6, but now the irradiation flux $F_{irr}$ is substituted by the stellar irradiation flux plus the ring flux, i.e.,

\be
F_{irr} = (1 - \Omega_A) F_{irr}^* + F_A, 
\en

\noindent
where $\Omega_A$ is the fraction of the solid angle subtended by the star, occupied by the ring (see the Appendix), and $F_{irr}^*$ is the irradiation flux from the star (see KH).

The cosine of the angle of incidence is estimated  as:

\be
\mu_0 = \Biggl \{ {\biggl ({1 \over \mu_*} T_*^4 (1-\Omega_A) + {1 \over \mu_A} T_A^4 \Omega_A \biggr ) \over \biggl ( T_*^4 (1-\Omega_A) +  T_A^4 \Omega_A \biggr ) } \Biggr \}^{-1}, 
\en

\noindent
where $\cos^{-1}(\mu_A)$ is the averaged angle between the 
 direction of the ring flux  and the normal to the disk surface (see the Appendix), and $\cos^{-1}(\mu_*)$ is the mean angle between 
the stellar radiative flux and the normal to the disk surface.
The mean Planck mean opacity, which describe the interaction between 
the disk material and the 
incident radiation field, is written as:

\be
\kappa_P^{irr} = \biggl \{ {\biggl (\kappa_P^* T_*^4 (1-\Omega_A) + \kappa_P^A T_A^4 \Omega_A \biggr ) \over \biggl ( T_*^4 (1-\Omega_A) +  T_A^4 \Omega_A \biggr ) } \biggr \}, 
\en

\noindent
and the other mean opacities $\chi^{irr}$ and $\chi_R^{irr}$ are calculated
with the same type of average.
 The mean ``ring'' opacities $\kappa_P^A$, $\chi_P^A$ and $\chi_R^A$,
 are the mean opacities calculated using the Planck function evaluated at $T_A$ as the weighting function (see eq. [\ref{eq_kaps}]).

 At every radius, the irradiation flux from the star is larger than 
the irradiation flux from the ring. For $R > 0.1 \ \AU$, the stellar flux 
is a factor of 10 larger than the ring flux.
 The ring irradiation increases the  disk photospheric temperature less than $1 \ \%$, in agreement with Kenyon \etal (1996). The temperature at the disk surface increases less than  $6 \ \%$.

\section{Summary and conclusions}

In this paper we discuss a
  model of the vertical structure of 
steady and thin accretion disks, irradiated by their central stars,
 that includes transport of energy by radiation, 
convection and turbulent flux.  The radiative transport is treated 
without the diffusion approximation: The optically thin and thick 
regions of the disk are calculated with the same equations, without an {\it a 
priori} separation.
We discuss the physical properties of a disk model 
with 
$\Mdot=10^{-8} \ \MSUNYR$ and $\alpha=0.01$, 
 around a low mass star with 
$M_*=0.5 \ \MSUN$, $R_*= 2 \ \RSUN$ and $T_*=4000 \ \K$, which 
 represents a typical 
T Tauri disk. The effect of other input parameters 
on the structure and physical properties  of disks will 
be reported in  paper II.

From this study, we
 conclude that the stellar irradiation is the main heating source 
of the outer regions ($R \gtrsim 1-2 \ \AU$)  in a typical 
T Tauri accretion disk, and has to be taken into account 
to calculate its 
physical properties,  
 such as:
 temperature, surface density, mass, height, timelife and gravitational 
stability. For instance,  
the photospheric and midplane temperature distributions of an irradiated disk 
are flatter than those of a non irradiated viscous disk. 
The former case has a smaller surface density and, therefore, a smaller total mass than  
 a non irradiated viscous disk. Moreover, irradiation tends to stabilize the 
disk against gravitational perturbations, 
increasing the Toomre's parameter above one.
Therefore,
 if some of the disk properties are inferred from observations,  
they can be related 
 to 
the disk parameters $\Mdot$ and $\alpha$, 
 through a 
detailed model, that must include the stellar irradiation.

 Stellar irradiation produces a temperature inversion at 
the disk atmosphere, 
 which, as shown by CPMD,  has observational consequences.
The sensitivity of the disk upper atmosphere temperature 
to the  mean opacity,    
suggests that it is important  to make a detailed  ``model 
atmosphere'' of the disk,  solving 
 the radiative transfer equation at different 
wavelengths and directions. Moreover, we have assumed that gas an dust are well 
mixed 
and  thermally coupled. HST observations of
HH30 (\cite{burrows96})  indicate that the dust scatters light 
from
the central object and producing the observed image at 
scale heights  $H/R \le 0.1$, supporting the assumption that gas and dust are well mixed. 
Nevertheless, in order to evaluate both assumptions it 
seems  necessary to calculate the disk structure considering gas and dust 
as separate ingredients with their particular dynamical and thermal  behavior.
 This is a problem outside the scope of this paper, but we want to emphasize 
the importance of both assumptions.  
On one hand, the gas is heated by viscous dissipation while dust is heated by stellar 
irradiation. If the collisional and radiative 
interaction between gas and dust 
is efficient enough, both can be characterized by the same  
 temperature. On the other hand, if gas and dust are 
well mixed by turbulent motions, 
they have the same spatial distribution. 
In any case, given both assumptions, the height of the 
disk can be calculated self-consistently 
including the irradiation.
 A simple estimate of the difference in temperature between 
gas and dust, including only 
radiative and collisional heating of the gas particles by dust grains, shows 
that  the assumption of a unique temperature 
is good  below the disk photosphere. Nevertheless, the region where 
the stellar radiation is mainly absorbed has such low pressure that coupling 
is not very efficient. This can introduce an inconsistency in 
the atmospheric calculation discussed in the present paper,
 which has to be evaluated with a 
separate calculation of gas and dust temperatures. Also, as discussed by 
Chiang \& Goldreich (1997), 
to calculate molecular line emission it is important to take into account
the possibility that gas and dust 
have different temperatures in the upper atmosphere.
In this sense, the present model corresponds to a first approximation to 
the atmospheric structure, which has to be evaluated with 
a more detailed treatment of the gas and dust thermal and dynamical behavior.

The accretion ring on the stellar surface can irradiated 
the disk. 
For the reference model this ring has a 
luminosity around $0.02-0.04 \ \LSUN$ and a temperature around $T_A=10000 \ \K$ (from the UV excess of T Tauri stars). In this case, we found that the 
effect of this additional heating source on the disk structure and emission, 
is negligible.

Our formalism can include any physical model that provides the viscosity
$\nu$ as a function of {\it local} physical conditions in the disk.
For the particular model discussed in this paper, we have used the 
$\alpha$ prescription, assuming a constant value of $\alpha$ throughout
the disk, to evaluate $\nu$.
 Given the vertical structure and following Gammie (1996), we have found 
the region where 
the ionization fraction (including thermal and energetic particles ionization) 
corresponds to a magnetic Reynolds number of $Re_{M}<1$, for a disk model 
with constant $\alpha$. In the region where $Re_{M} < 1$ the magnetic 
instability proposed by Balbus \& Hawley (1991) cannot work. If the source of 
turbulence is 
this magnetic instability,  then there cannot be viscous dissipation 
where $Re_{M} < 1$, i.e., for $0.2 < R < 4 \ \AU$,  
 and this region will represent the dead zone 
of Gammie's model. 
 In particular, Gammie proposes a ``layered accretion'' throughout this
dead zone to alleviate the problem of the non-steady behavior of
accretion disks.

\begin{figure}
\plotone{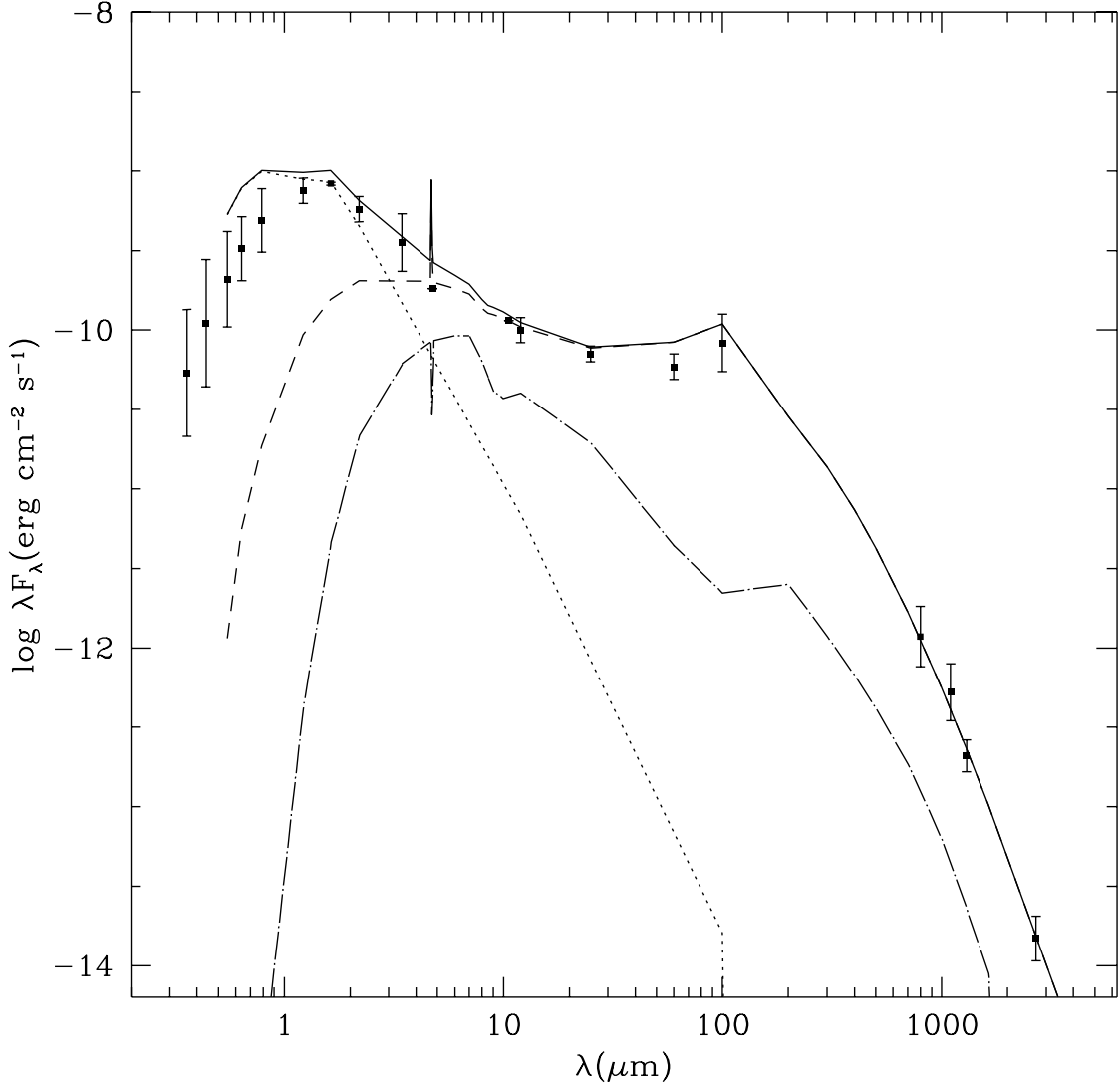}
\caption {SED of the reference model pole-on, compared with AA Tau observed SED. 
  The system's SED (star+disk) is shown with a solid line, 
the central star SED, taken from Bruzual \& Charlot (1993), is plotted with a dotted line,  
  and the disk SED is plotted with a dashed line.
Also, the SED of a purely viscous disk with the same $\Mdot$, $\alpha$ and 
central star, is shown with a dot-dashed line. 
The observational points (squares) were taken from Adams \etal (1990), 
  Beckwith \etal (1990), Beckwith \& Sargent (1991), Weaver \& Jones (1992), 
Kenyon \& Hartmann (1995), Dutrey \etal (1996). The error bars 
are related to the variability of the object for  $\lambda < 12 \ \mu m$, 
and represent the reported errors for 
 $\lambda \ge 12 \ \mu$.}
\label{fig_sed}
\end{figure}

Finally, the predicted SED for the reference model 
is in agreement with  the SED of AA Tau, for which preliminary 
results from 
Gullbring \etal (1997) are: $\Mdot=3.8 \times 10^{-9} \ \MSUNYR$, $M_*=0.53 \ 
\MSUN$, $R_*=1.74 \ \RSUN$ and $T_*=4048 \ \K$.  
Both SEDs are shown in Figure \ref{fig_sed}. 
The wavelengths at which the SED is plotted, are thought to 
sample the continuum spectrum, and not to 
resolve lines and bands.  Nevertheless, there are some 
apparent features in the irradiated disk SED, like 
the  CO fundamental tone in emission around 4.3 $\mu {\rm m}$ and an IR band of water vapor around 100 $\mu {\rm m}$. It is important to emphasize that these bands 
in emission are due to the atmospheric temperature inversion, 
and the assumption that gas and dust are thermally coupled. If this last 
assumption is not valid at the upper disk atmosphere, the details of the 
gas spectral features will change respect to what is shown in Figure \ref{fig_sed} (see \cite{CG97}).
Details of the disk SEDs, calculated for different inclination angles, will be discussed in a forthcoming paper.

\acknowledgments
We are grateful to  Javier Ballesteros, Lee Hartmann, 
Alejandro Raga and Salvador Curiel 
 for helpful discussions.
  This work was supported in part by Instituto de Astronom\'\i a, UNAM, M\'exico, DGAPA-UNAM, ConaCyT, 
 NAGW 2306 and NAG5-4282.

\clearpage
\appendix
\section{Irradiation of the disk by a hot ring on the stellar surface}

\begin{figure}
\plotone{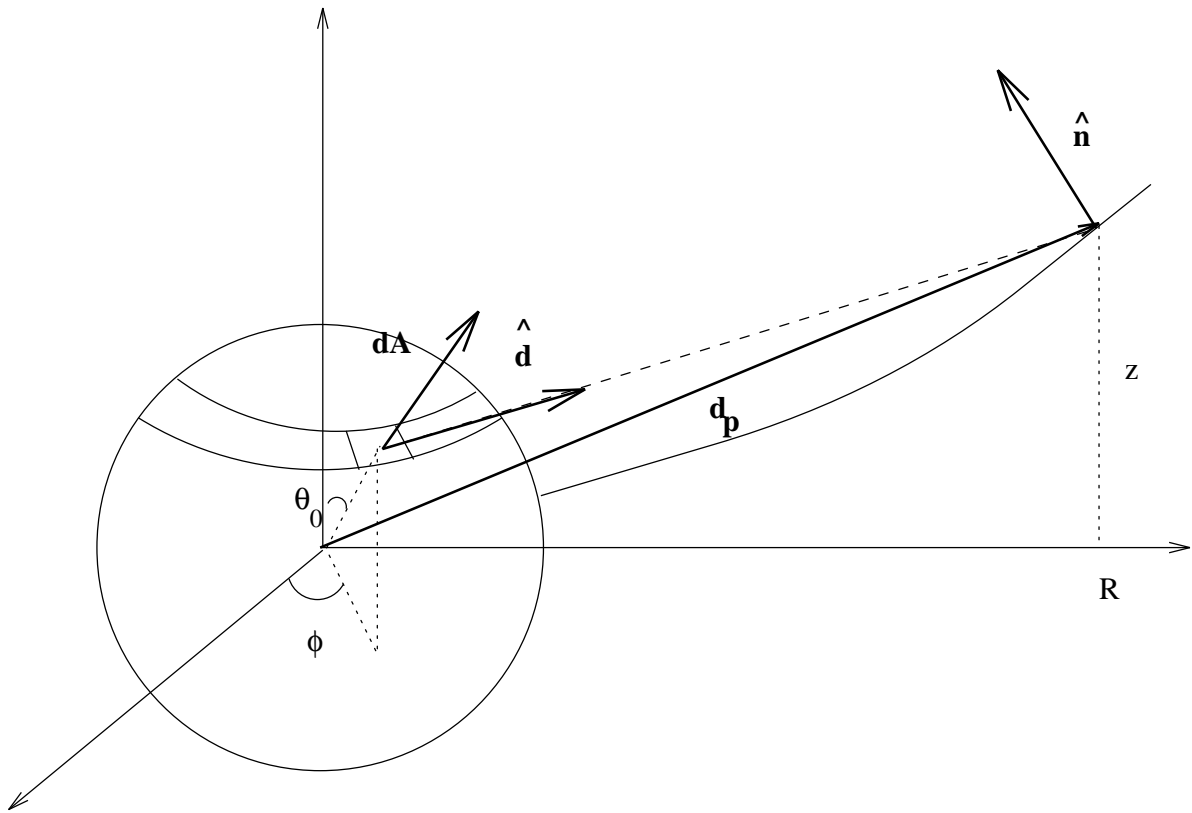}
\caption {Geometry used to describe the irradiation of the disk 
due to a hot ring  at the stellar surface.}
\label{fig_annulus}
\end{figure}

In this appendix we describe the calculation of the flux intercepted by the disk, emergent from a hot annulus on the stellar surface. This annulus would correspond to  
the accretion shock between the material flowing through the stellar magnetosphere 
and the surface of the star.
 Figure \ref{fig_annulus}  shows the geometry of this problem. 
The coordinate system and angles are defined in this figure.

A surface  element of the hot annulus is given by: 

\be
d \vec{A}=R_*^2 \ d\Omega \ (\sin \theta \cos \phi \hat{x} + 
\sin \theta \ \sin \phi \hat{y} + \cos \theta \hat{z})
\en

\noindent
where $d\Omega=\sin \theta d \theta d\phi$ is the solid angle of the surface element as seen from the center of the star.
 The same surface element, seen from a point $P$ on the disk surface, 
has an apparent area $dA_p$, given by the projection of $d \vec{A}$ on $\hat{d}$, where  
$\vec{d}=\vec{d_p} - \vec{r}$, $\vec{d_p}$
is the vector describing the position of  $P$,
 and $\vec{r}$ is the position vector of the annulus element, both given by:

\be
\vec{r}=R_*(\sin \theta \cos \phi \hat{x} + 
\sin \theta \ \sin \phi \hat{y} + \cos \theta \hat{z})
\en

\be
\vec{d_p}=R \hat{y}+ z_\infty \hat{z}
\en

Thus, the unit vector $\hat {d}$  can be written as:

\be
\hat{d}={\vec{d_p} - \vec{r} \over |\vec{d_p} - \vec{r}|}=
{-R_* \sin \theta \cos \phi \hat{x}+(R-R_* \sin \theta \sin \phi) \hat{y}+
(z_\infty-R_* \cos \theta) \hat{z} \over [R_*^2+R^2+z_\infty^2-2RR_* \sin \theta \sin \phi 
-2 z_\infty R_* \cos \theta]^{1/2}}
\en

The annulus  element,  as seen from point  $P$, has a 
 solid angle given by:

\be
d \Omega_p = {dA_p \over {d}^2}= {d \Omega R_*^2 [R \sin \theta \sin \phi 
-R_* + z_\infty \cos \theta] \over [R_*^2+R^2+z_\infty^2-2 R R_* \sin \theta \sin \phi - 2 z_\infty R_* \cos \theta]^{3/2}}
\en

The mean intensity of the annulus radiation field at point $P$ is 
given by:

\be
J_\nu= {1 \over 4 \pi} \int I_\nu^{A} d \Omega_p
\en

The flux intercepted by the disk is:

\be
F_A= \int_0^\infty \int I_\nu^{A} \mu d \Omega_p d\nu= {\sigma_R T_A^4 \over \pi} \int \mu d \Omega_p
\label{eq_fluxhot}
\en

\noindent
where $T_A$ is the ring effective temperature and $\mu$ is the cosine of the angle between the vector $\hat{d}$, which characterizes the direction of incidence of the radiation, and the vector $\hat{n}$, normal to the disk surface. The disk normal vector can be written as:

\be
\hat{n}={-(dz_\infty/dR) \hat{y}+\hat{z} \over (1+(dz_\infty/dR)^2)^{1/2}}
\en

Thus,
\be
\mu={-\hat{d} \cdot \hat{n}}={ [(R-R_* \sin \theta \sin \phi) dz_\infty/dR - z_\infty+R_* \cos \theta]  \over  (1+(dz_\infty/dR)^2)^{1/2} [R_*^2+R^2+z_\infty^2-2 R R_* \sin \theta \sin \phi - 2 z_\infty R_* \cos \theta]^{1/2}}
\label{eq_muhot}
\en

Finally, substituting equation (\ref{eq_muhot}) in (\ref{eq_fluxhot}), 
the flux of the annulus intercepted by the disk can be calculated.
The averaged value of the  cosine of the angle between the incident 
radiation and the disk normal can be calculated as:

\be
\mu_{A} = -{\int \hat{d} \cdot \hat{n} d \Omega_p \over 
\int d \Omega_p}
\en

For simplicity, we assume that the annulus is at $\theta=\theta_0$ and it 
is thin enough to approximate the integrals over $\theta$ 
by the integrand evaluated at $\theta_0$ times $\delta \theta_0$.
For $R < (R_*-z_\infty \cos \theta_0)/\sin \theta_0$, the annulus is no {\it seen} by the disk.
The integral over $\phi$ is symmetric respect to $\phi=\pi/2$, then 
we make it from $\phi_{min}$ to $\phi_{max}$, and multiply 
the result by 2. The limits of this integration are given by:

\be
 \phi_{min}=\sin^{-1} {R_* -z_\infty \cos \theta_0 \over R \sin \theta_0}
\en

\noindent
if $|R_* -z_\infty  \cos \theta_0| \le  |R \sin \theta_0|$, otherwise  
$\phi_{min}=-\pi/2$, and $\phi_{max}=\pi/2.$

\clearpage

\end{document}